\documentclass{article}

\usepackage{arxiv}

\usepackage[utf8]{inputenc} 
\usepackage[T1]{fontenc}    
\usepackage{hyperref}       
\usepackage{url}            
\usepackage{booktabs}       
\usepackage{amsfonts}       
\usepackage{nicefrac}       
\usepackage{microtype}      
\usepackage{lipsum}
\usepackage{graphicx}
\usepackage{amsmath}
\graphicspath{ {./images/} }

\title{Artificial Intelligence and Misinformation in Art: Can Vision Language Models Judge the Hand or the Machine Behind the Canvas?}

\author{
 Tarian Fu \\
  Nanjing University of Aeronautics and Astronautics\\
   \And
 Javier Conde \\
  Universidad Politécnica de Madrid\\
  \And
 Gonzalo Martínez \\
  Universidad Politécnica de Madrid \\
    \And
    Pedro Reviriego \\
  Universidad Politécnica de Madrid\\
   \And
 Elena Merino-Gómez \\
  Universidad de Valladolid\\
  \And
 Fernando Moral \\
  Universidad Antonio de Nebrija \\
}

\begin{document}
\maketitle

\begin{abstract}
The attribution of artworks in general and of paintings in particular has always been an issue in art. The advent of powerful artificial intelligence models that can generate and analyze images creates new challenges for painting attribution. On the one hand, AI models can create images that mimic the style of a painter, which can be incorrectly attributed, for example, by other AI models. On the other hand, AI models may not be able to correctly identify the artist for real paintings, inducing users to incorrectly attribute paintings. In this paper, both problems are experimentally studied using state-of-the-art AI models for image generation and analysis on a large dataset with close to 40,000 paintings from 128 artists. The results show that vision language models have limited capabilities to: 1) perform canvas attribution and 2) to identify AI generated images. As users increasingly rely on queries to AI models to get information, these results show the need to improve the capabilities of VLMs to reliably perform artist attribution and detection of AI generated images to prevent the spread of incorrect information.      
\end{abstract}







\section{Introduction}

\begin{figure}[h]
    \centering
    \includegraphics[width=0.35\textwidth]{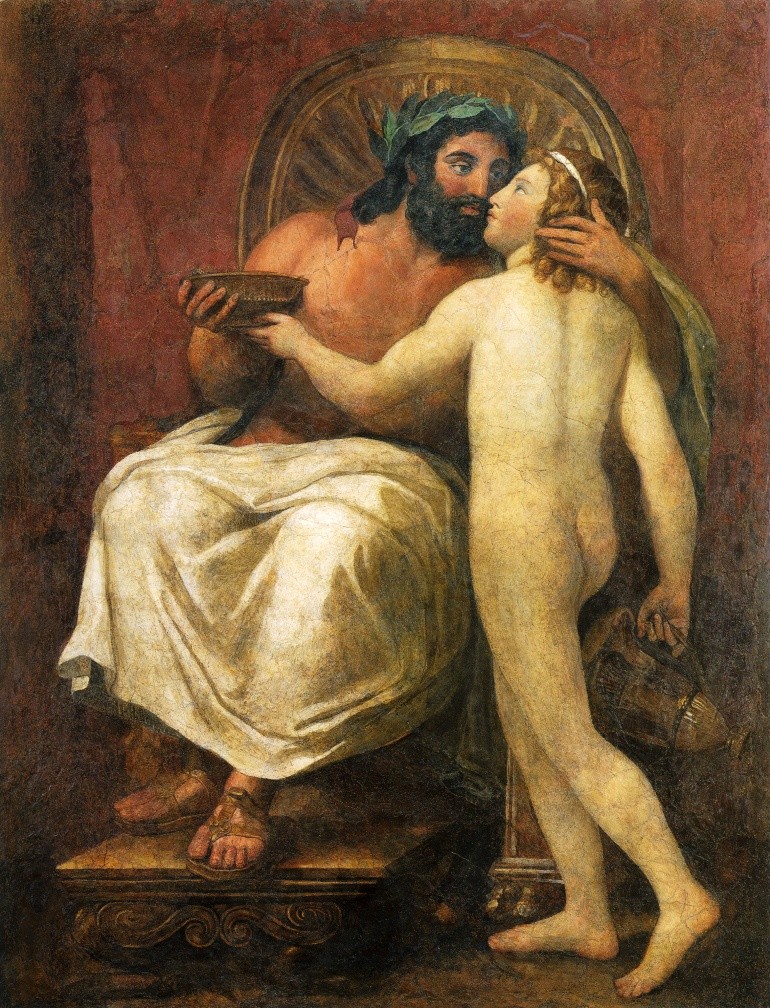}
    \caption{Anton Raphael Mengs, \textit{Jupiter kissing Ganymede}, c. 1755.}
    \label{fig:mengs}
\end{figure}

The attribution of works has always been a fundamental issue in art history and the cause of many disputes. Notorious is the fake ancient Roman fresco with which the painter Anton Raphael Mengs (1728–1779) deceived Johann Joachim Winckelmann, the eminent art historian and theorist of Neoclassicism. Created around 1755, the scene of Jupiter kissing Ganymede (see Figure \ref{fig:mengs}) was intended as a stylistic homage to antiquity, but was presented as authentic. Winckelmann, convinced of its antiquity, praised it as a rare survival of Greco-Roman painting, thereby exposing the vulnerability of even the most refined connoisseurly judgment to forgeries when driven by idealistic expectations about the classical past \cite{honour1977}. Mengs later confessed the deception, underscoring the subjective limits of connoisseurship in the pre-scientific era of art historical evaluation.

More than a century later, the modern art world was shaken by the infamous forgeries of Han van Meegeren, who successfully passed off multiple paintings in the style of Johannes Vermeer, including "The Supper at Emmaus" (1937), as authentic 17th-century Dutch masterpieces. Acclaimed by leading Vermeer expert Abraham Bredius and purchased by Dutch museums, van Meegeren's forgeries exposed both a desire to "complete" Vermeer’s sparse oeuvre and the absence of rigorous technical analysis at the time. His eventual confession in 1945, delivered during a trial in which he was accused of selling Dutch cultural property to the Nazis, turned him from traitor to national hero, and marked a turning point in forensic art authentication \cite{dutton1979}. Modern pigment analysis, particularly the detection of synthetic ultramarine and phenol-formaldehyde resins, later proved vital in debunking his paintings' historical authenticity \cite{coremans1949}.

Perhaps even more challenging is the question of authorship in the oeuvre of Rembrandt van Rijn. Unlike Vermeer, Rembrandt worked in a bustling studio environment and often encouraged his students to emulate his style closely. Complicating matters further is his inconsistent signing practice (sometimes abbreviated, sometimes fully written, and often absent) rendering signatures an unreliable tool for authentication. The launch of the Rembrandt Research Project (RRP) in 1968 brought systematic scrutiny to Rembrandt attribution. Under the leadership of Ernst van de Wetering, the RRP employed a combination of connoisseurship, archival research, and advanced imaging techniques (including X-radiography and dendrochronology) to reassess the authenticity of hundreds of works. As a result, many previously accepted paintings were downgraded, while others were reattributed to Rembrandt after long exclusion \cite{wetering2004, wetering2015}.

A wide range of Artificial Intelligence (AI)-driven techniques, such as the use of information-based processing \cite{InformationBasedAttribution}, deep transfer learning \cite{Raphael}, and surface scanning of the canvas \cite{surface}, have been explored for artist attribution \cite{Rembrandt},\cite{MadonnaAttribution}. Those studies focus on specific artists or techniques and the tools developed are not available to the general public.

The advent of powerful vision language models capable of advanced image analysis \cite{VLMs} offers additional tools for artist attribution \cite{GalleryGPT}. These models have been trained with billions of images and can answer sophisticated questions on almost any kind of image. In fact, any user can upload an image of a painting and ask the model about it. This can be an issue if their responses are not correct as they may create confusion or even disinformation on users who tend to query and trust AI models \cite{sun2024trusting}. Therefore, it is of interest to evaluate the capabilities of state-of-the-art vision language models to perform artist attribution.

The impact of generative AI on artist attribution does not end with vision language models, the development of powerful text-to-image models \cite{diffusion_latent} enables users to create images imitating a given painter or style \cite{asperti2025critical} or even to modify real paintings \cite{RIP}. This can lead to additional confusion for artist attribution by having AI-generated images attributed to artists. An interesting twist is when vision language models are presented with an AI-generated painting imitating an artist. Would the model incorrectly attribute the image to the painter or would it recognize that it was generated by another AI tool? Exploring this issue is also of interest, as more and more AI-generated content populates the Internet.

In this paper, we present an extensive experimental evaluation of the capabilities of a set of relevant vision language models when performing artist attribution on a dataset with close to 40,000 images of paintings from 128 artists, and AI-generated imitations of those paintings. The main contributions of this work are:

\begin{enumerate}
    \item To evaluate and analyze the capabilities for artist attribution of real paintings of vision language models at scale.
    \item To evaluate and analyze the capabilities for artist attribution of AI-generated paintings of vision language models at scale.
    \item To make available a dataset of AI-generated descriptions of real paintings and Web interface for visualization\footnote{The code and raw data are available at \url{https://github.com/aMa2210/WikiArt_VLM_Web}. In addition, a website is also available to interactively visualize the results \url{https://ama2210.github.io/WikiArt_VLM_Web/}.  }.
    \item To make available a dataset of AI-generated images that mimic the style of the artists.
    \item To discuss the implications of vision language models artist attribution performance as generative AI adoption becomes widespread.   
\end{enumerate}

The rest of the paper is organized as follows, section \ref{relatedwork} discusses related works on the use of AI for artists attribution and image generation. The methodology used in the evaluation is presented in section \ref{methodolody} and the results as well as the limitations of the study are discussed in section \ref{results}. The paper ends with the conclusion in section \ref{conclusion}.

\section{Related work}
\label{relatedwork}

The use of image processing and machine learning models for the identification of artists has been explored for decades \cite{johnson2008image,saleh2015large}. Initially, simple models such as support vector machines (SVMs) operating on different features extracted from the painting were proposed. The rapid development of Convolutional Neural Networks (CNNs) for image processing that achieve excellent performance in several tasks \cite{johnson2008image} led to the use of CNNs for the identification of artists \cite{tan2016ceci} and more recently, to the use of transformers \cite{ArtVisionTransformers}. All of these models and tools are specialized and not widely available to users.

The development of Vision Language Models (VLMs) that can combine text and image \cite{radford2021learning} has been a revolution \cite{VLMs}. VLMs can answer almost any question about an image and are available to users in applications such as ChatGPT that are used by billions of people every day. The use of VLMs has been proposed, for example, to explain artworks \cite{hayashi2024towards}. Today, any user can upload an image of a painting and ask the VLM for the artist who drew the painting. In fact, VLMs have been evaluated to identify painting styles and have been shown to achieve lower accuracy than specific tools \cite{strafforello2024have}. This is worrying as users are increasingly dependent on VLM based applications and assistants to access information. However, to the best of our knowledge, no large-scale evaluation of VLM performance when used for artist identification has been reported in the literature.  

Another area that has experienced impressive progress in recent years is image generation from text prompts \cite{yang2025texttoimage}. Again, there are many publicly available models, such as Stable Diffusion \cite{SD} that can generate all sorts of images. These are incorporated into tools so that users can easily create images at will, for example, imitating a given artist \cite{imitatingSD}. This adds another dimension to the identification of artists, as now there is a need to also detect and discriminate images created by AI models. Although specific models can be designed to detect AI-generated images \cite{li2025detecting}, users are more likely to ask general-purpose VLMs for an answer. Therefore, there is further interest in understanding whether VLMs can identify AI-generated images and not attribute them to painters even when they mimic their style. Again, to the best of our knowledge, no large-scale evaluation of VLM performance when used for artist identification has been reported in the literature when run on AI-generated images.

\section{Methodology}
\label{methodolody}

To evaluate the performance of vision language models in artist attribution, we have to select a relevant dataset of images and models to evaluate. In the case of AI-generated images, no such dataset was found at the time of writing this paper, and therefore we created it as part of this work. We also need to define the procedure used for the evaluation as well as the metrics used to analyze the results. The following subsections discuss each of these issues in detail.

\subsection{Real paintings dataset}

To perform an evaluation at scale, we have selected the $WikiArt$ dataset\footnote{\url{https://huggingface.co/datasets/huggan/wikiart}} that contains paintings by 128 artists covering 10 genres and 27 styles. Each image in the dataset has the artist, genre, and style as metadata. Images with "unknown" artists are not considered leaving 39,530 images. This dataset provides a sufficient number of artists and paintings and is publicly available which facilitates reproducing or extending our research.

\subsection{AI-generated paintings dataset}

\begin{figure}[htbp]
    \centering
    \includegraphics[width=0.7\linewidth]{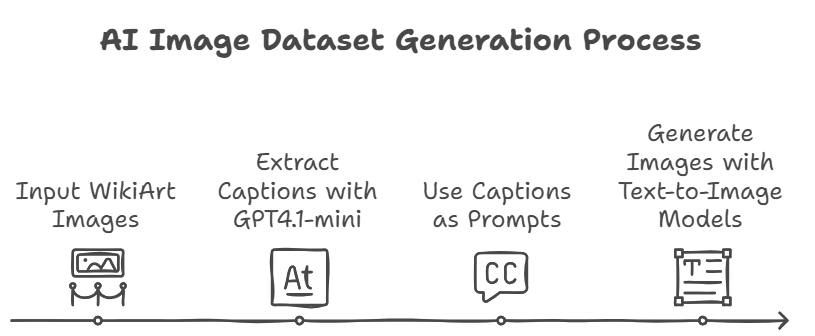}
    \caption{Process to generate the AI painting imitations}
    \label{fig:dataset-generation}
\end{figure}

In the case of paintings generated by AI, it was not possible to find a suitable dataset to perform an evaluation at scale. Therefore, we decided to create it as part of this work. To do it, we first extracted the caption from the 39,530 WikiArt images using GPT4.1-mini\footnote{The version used was gpt-4.1-mini-2025-04-14.}. Then the prompts were used to generate images with three text-to-image models: Stable Diffusion\footnote{The version used was stable-diffusion-3.5-large.}, Flux\footnote{The version used was FLUX.1-dev.}, and F-Lite\footnote{The version used was 
F-Lite.}. The general process is illustrated in Figure \ref{fig:dataset-generation}. The prompts are publicly available to facilitate the creation of data sets with other text-to-image models. The prompt used to generate the images has the following structure: ``\textit{Produce an image that closely resembles a painting by <correct painter>, but is not an exact copy of his works: <caption of the real painting>}'' The images created with Stable Diffusion, Flux and F-Lite are also in the same repository so that they can be reused in other works\footnote{The prompts and images are available at \url{https://github.com/aMa2210/WikiArt_VLM}}. An advantage of the method used to generate the images is that the real and AI-generated datasets are homogeneous in terms of number and type of images, which makes comparisons between datasets more meaningful.

\subsection{Vision language models}

A group of five open-weight vision language models from different companies has been selected for evaluation. These models can be run locally in off-the-shelf GPUs. The set is completed with a proprietary model from Open AI. The six models evaluated are:

\begin{enumerate}
    \item GPT4.1-mini: a model from OpenAI\footnote{The version of the model is gpt-4.1-mini-2025-04-14}. 
    \item Gemma3-12B: a model from Google\footnote{The model is \url{https://huggingface.co/google/gemma-3-12b-it}}.
    \item LLaMa3.2-11B a model from Meta\footnote{The model is \url{https://huggingface.co/meta-llama/Llama-3.2-11B-Vision-Instruct}}. 
    \item Phi-4-5.6B a model from Microsoft\footnote{The model is \url{https://huggingface.co/microsoft/Phi-4-multimodal-instruct}}.
    \item QwenVL-2.5-7B a model from Alibaba\footnote{The model is \url{https://huggingface.co/Qwen/Qwen2.5-VL-7B-Instruct}}.
    \item Pixtral-12B a model from Mistral\footnote{The model is \url{https://huggingface.co/mistral-community/pixtral-12b}}. 
\end{enumerate}

This group of models provides a sample of vision language models that is sufficient to extract relevant conclusions while keeping the computational effort and cost manageable. 

\subsection{Evaluation procedure}

As we want to conduct an evaluation at scale on tens of thousands of paintings on several models, the process has to be automated to be manageable. This poses some limitations on how to ask the models for the author of a painting. If we ask an open question, the model may reply with a reasoning from which it may be hard to get the name of the artist. We can ask the model to just give the name of the author or provide the explanation and then end with the name in a given format, for example in brackets. However, the model can produce the name of an artist in different ways, for example give just the surname or the full name. This makes the parsing of the responses complex and error prone. To avoid this problem, we have used a simple prompt: 

Prompt-1 correct artist: \textit{"Is this a real painting from <correct painter>? Please answer only {yes} or {no}"}

Taking the \textit{"correct painter"} from the metadata in the WikiArt dataset. 

This strategy makes the processing simple to automate but has a potential problem as a model that always answers \textit{yes} will get 100\% accuracy. To ensure that models can discriminate paintings, we use a second prompt:  

Prompt-2 incorrect artist: \textit{"Is this a real painting from <incorrect painter>? Please answer only {yes} or {no}"}

Asking if the painting corresponds to another painter, different from the author selected randomly from the remaining 127 artists. 

\begin{figure}[htbp]
    \centering
    \includegraphics[width=0.7\linewidth]{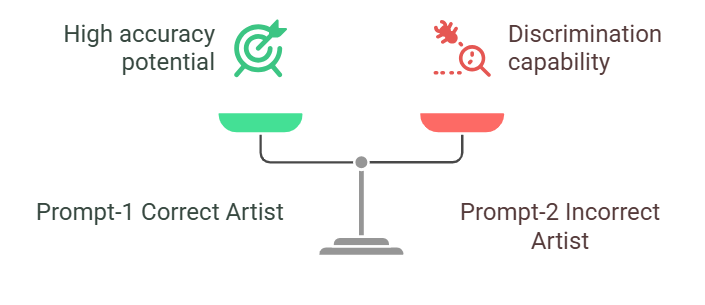}
    \caption{Process to evaluate the models}
    \label{fig:eval-process}
\end{figure}

The dataset is run twice, once with each prompt, to assess the model's capability to identify the author and also to detect that it was not painted by other artists. The process and rationale are illustrated in Figure \ref{fig:eval-process}. We run the same two promts with every AI-generated image too. In this case, the correct answer is \textit{no} for both prompts, but the difference in the attribution rates for both prompts can also be informative. 

\subsection{Evaluation Metrics}

To evaluate the performance of vision language models, we will use two individual metrics $C1,C2$ that are the correctness of the responses for Prompt-1 and Prompt-2, respectively normalized to the random guess value of 50\% for a two-response question as follows: 


\begin{equation}
C = \frac{\text{Percentage of correct responses} - 50}{50}
\end{equation}

As each of those individual metrics provides information only on one aspect of the performance, we also propose to use a combined metric, the arithmetic mean $A_M$ calculated as:


\begin{equation}
A_M = \frac{C1 + C2}{2}
\end{equation}




For real paintings, ideally, both $C1,C2$ will be close to one, and the arithmetic mean will only approach one when both are close to one. 

For AI-generated paintings, ideally, both $C1,C2$ will also be close to one, taking into account that now in both cases the correct answer is \textit{no}. Therefore, the arithmetic mean which captures the ability of the models to identify that both AI-generated images mimicking the style of the painter (Prompt-1) or of a different painter (Prompt-2) are not attributed to any painter is also a relevant metric.

\section{Results and analysis}
\label{results}

This section presents the results of the experimental evaluation. First, we discuss the results of running Prompt-1 and Prompt-2 on images of real paintings. Then the results of the evaluation on AI-generated images that mimic paintings are presented and the limitations of the study discussed. The section ends with an analysis and discussion of the results.

\subsection{Real paintings}

\begin{figure}[htbp]
    \centering
    \includegraphics[width=0.8\linewidth]{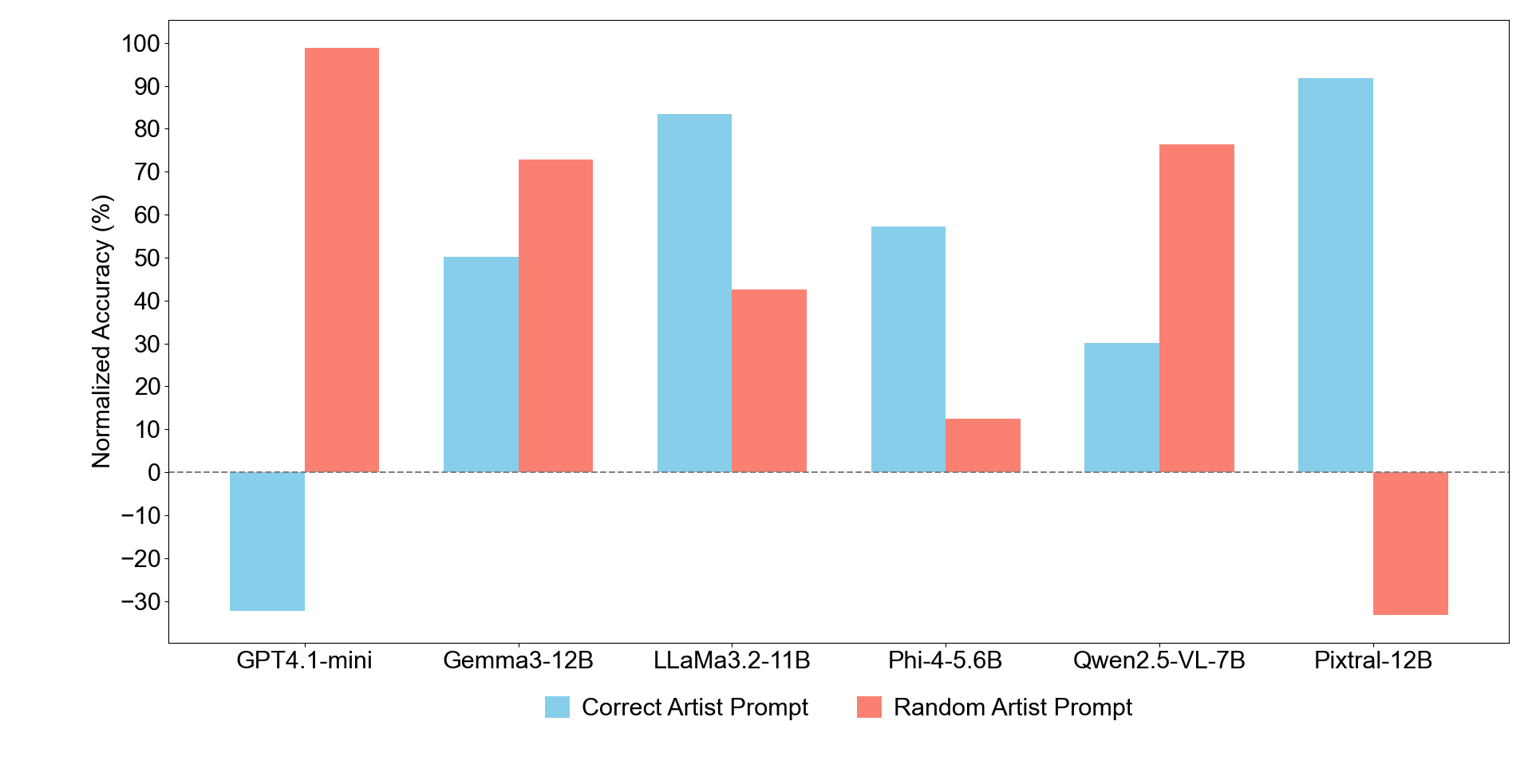}
    \caption{Average $C1$ and $C2$ scores on all painters for the VLMs considered on WikiArt images of real paintings}
    \label{fig:C1_C2}
\end{figure}

The average results for all artists for $C1,C2$ are shown in Figure \ref{fig:C1_C2} per model. It can be seen that the results vary significantly between models. GPT4.1-mini does not attribute paintings neither to the real author ($C1$) nor to a random painter ($C2$). Instead, Pixtral-12B answers correctly most of the time when the painter is the real one and fails when the painter is a random one. They are examples of a conservative model that tends not to attribute paintings (GPT4.1-mini) and a more aggressive model that tends to attribute the painting to the suggested painter (Pixtral-12B). Both are undesired behaviors being aggressiveness potentially more dangerous from a misinformation perspective. The rest of the models present more even values of $C1,C2$ with Gemma3-12B and LLaMa3.2-11B achieving more than 40\% normalized correct answers in both $C1,C2$. 

\begin{figure}[htbp]
    \centering
    \includegraphics[width=0.8\linewidth]{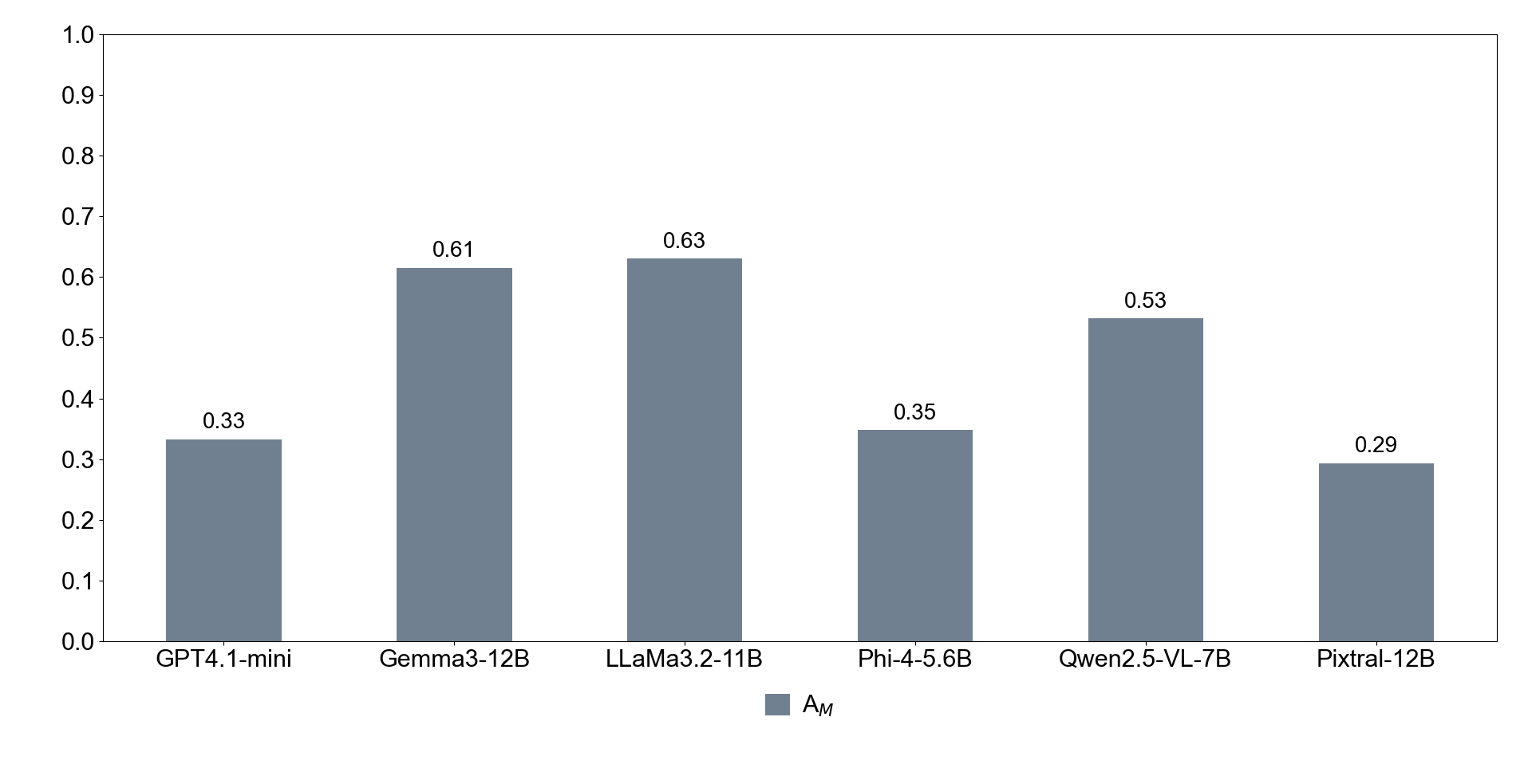}
    \caption{Average $A_M$ scores on all painters for the VLMs considered on WikiArt images of real paintings}
    \label{fig:Am_Hm}
\end{figure}

The combined metric $A_M$ is shown per model in Figure \ref{fig:Am_Hm}. The results show that again Gemma3-12B and LLaMa3.2-11B are the best performing models.

\begin{figure*}[htbp]
    \centering
    \includegraphics[width=\linewidth]{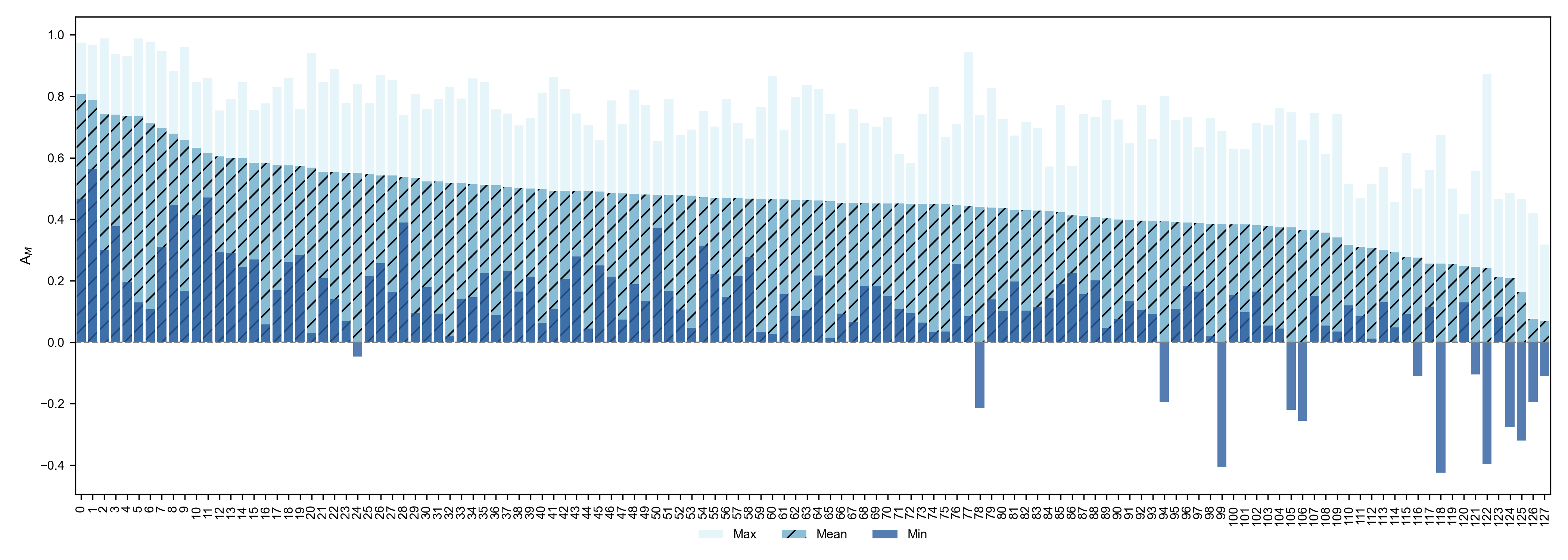}
    \caption{Painters ordered by average $A_M$ scores for the VLMs considered on WikiArt images of real paintings. The maximum and minimum $A_M$ scores across the VLMs are also shown.}
    \label{fig:artists_AM}
\end{figure*}

The results per painter are shown in Figure \ref{fig:artists_AM} with the correspondence of numbers with artists in Table \ref{tab:artist_list}. It can be seen that there are large differences in performance between painters. The painter with best results, Utagawa Kuniyoshi, has over 80\% normalized average accuracy, while the worst, M.C. Escher, has almost 0\% normalized average accuracy. The popularity of the artist does not seem to help VLMs to recognize their paintings as Vincent Van Gogh or Salvador Dalí are among the bottom 10 models. It is also worth noting that universally known artworks, such as La Gioconda by Leonardo da Vinci or The Kiss by Gustav Klimt, are not recognized as authentic by any of the VLMs considered. 

In summary, the evaluation on a large set of painters shows that current VLMs have strong limitations when identifying the artist of real canvas and thus cannot be considered a reliable source of information.

\begin{table*}[htbp]
  \centering
  \tiny
  \begin{tabular}{rlrlrlrl}
    \toprule
    No. & Artist & No. & Artist & No. & Artist & No. & Artist \\
    \midrule
    1 & Utagawa Kuniyoshi & 33 & Aubrey Beardsley & 65 & Edgar Degas & 97 & Valentin Serov \\
    2 & Lucas Cranach the Elder & 34 & Theo Van Rysselberghe & 66 & El Greco & 98 & Giovanni Boldini \\
    3 & Fra Angelico & 35 & Nikolay Bogdanov Belsky & 67 & Dante Gabriel Rossetti & 99 & Andy Warhol \\
    4 & Juan Gris & 36 & Felix Vallotton & 68 & Nicholas Roerich & 100 & Francisco Goya \\
    5 & Pietro Perugino & 37 & Camille Pissarro & 69 & Martiros Saryan & 101 & James Tissot \\
    6 & Antoine Blanchard & 38 & Raoul Dufy & 70 & Ivan Kramskoy & 102 & Odilon Redon \\
    7 & Edouard Cortes & 39 & Koloman Moser & 71 & Ilya Repin & 103 & Jacek Malczewski \\
    8 & Hans Memling & 40 & Berthe Morisot & 72 & Ferdinand Hodler & 104 & Leonardo Da Vinci \\
    9 & Jacob Jordaens & 41 & Alfred Sisley & 73 & Ernst Ludwig Kirchner & 105 & John Singer Sargent \\
    10 & Albrecht Durer & 42 & Maurice Prendergast & 74 & Henri Martin & 106 & Pablo Picasso \\
    11 & Ivan Bilibin & 43 & Konstantin Makovsky & 75 & Gene Davis & 107 & Michelangelo \\
    12 & Hans Holbein the Younger & 44 & Henri Matisse & 76 & Camille Corot & 108 & Sir Lawrence Alma Tadema \\
    13 & Paolo Veronese & 45 & Sam Francis & 77 & Karl Bryullov & 109 & Isaac Levitan \\
    14 & Hieronymus Bosch & 46 & Titian & 78 & Rembrandt & 110 & Ivan Shishkin \\
    15 & Frans Hals & 47 & Vasily Perov & 79 & Canaletto & 111 & James Mcneill Whistler \\
    16 & Anthony Van Dyck & 48 & Aleksey Savrasov & 80 & Henri De Toulouse Lautrec & 112 & Vasily Vereshchagin \\
    17 & Gustave Loiseau & 49 & Peter Paul Rubens & 81 & William Merritt Chase & 113 & Arkhip Kuindzhi \\
    18 & Joshua Reynolds & 50 & Mstislav Dobuzhinsky & 82 & Mary Cassatt & 114 & Paul Gauguin \\
    19 & Tintoretto & 51 & Niko Pirosmani & 83 & Zinaida Serebriakova & 115 & Jan Matejko \\
    20 & Georges Braque & 52 & Raphael & 84 & Gustave Caillebotte & 116 & Georges Seurat \\
    21 & Gustave Dore & 53 & William Turner & 85 & Henri Fantin Latour & 117 & Lucian Freud \\
    22 & Pierre Auguste Renoir & 54 & John Henry Twachtman & 86 & Gustav Klimt & 118 & Viktor Vasnetsov \\
    23 & Katsushika Hokusai & 55 & Amedeo Modigliani & 87 & Thomas Gainsborough & 119 & Joaquín Sorolla \\
    24 & Eugene Boudin & 56 & Bartolome Esteban Murillo & 88 & Kuzma Petrov Vodkin & 120 & Edvard Munch \\
    25 & Fernando Botero & 57 & Orest Kiprensky & 89 & Konstantin Somov & 121 & Thomas Eakins \\
    26 & Joseph Wright & 58 & Edward Burne Jones & 90 & Childe Hassam & 122 & Mikhail Vrubel \\
    27 & Marc Chagall & 59 & Gustave Moreau & 91 & Vasily Polenov & 123 & Vincent Van Gogh \\
    28 & Pyotr Konchalovsky & 60 & Maxime Maufra & 92 & Ilya Mashkov & 124 & Vasily Surikov \\
    29 & Fernand Leger & 61 & David Burliuk & 93 & Mikalojus Ciurlionis & 125 & Edouard Manet \\
    30 & Ivan Aivazovsky & 62 & Boris Kustodiev & 94 & Pierre Bonnard & 126 & Eugene Delacroix \\
    31 & Paul Cezanne & 63 & Konstantin Korovin & 95 & Gustave Courbet & 127 & Salvador Dalí \\
    32 & Egon Schiele & 64 & Claude Monet & 96 & Raphael Kirchner & 128 & M.C. Escher \\
    \bottomrule
  \end{tabular}
  \caption{List of 128 Artists Sorted by Arithmetic Mean Normalized Accuracy on WikiArt images of real paintings}
  \label{tab:artist_list}
\end{table*}





\subsection{AI-generated paintings}

\begin{figure}[htbp]
    \centering
    \includegraphics[width=0.8\linewidth]{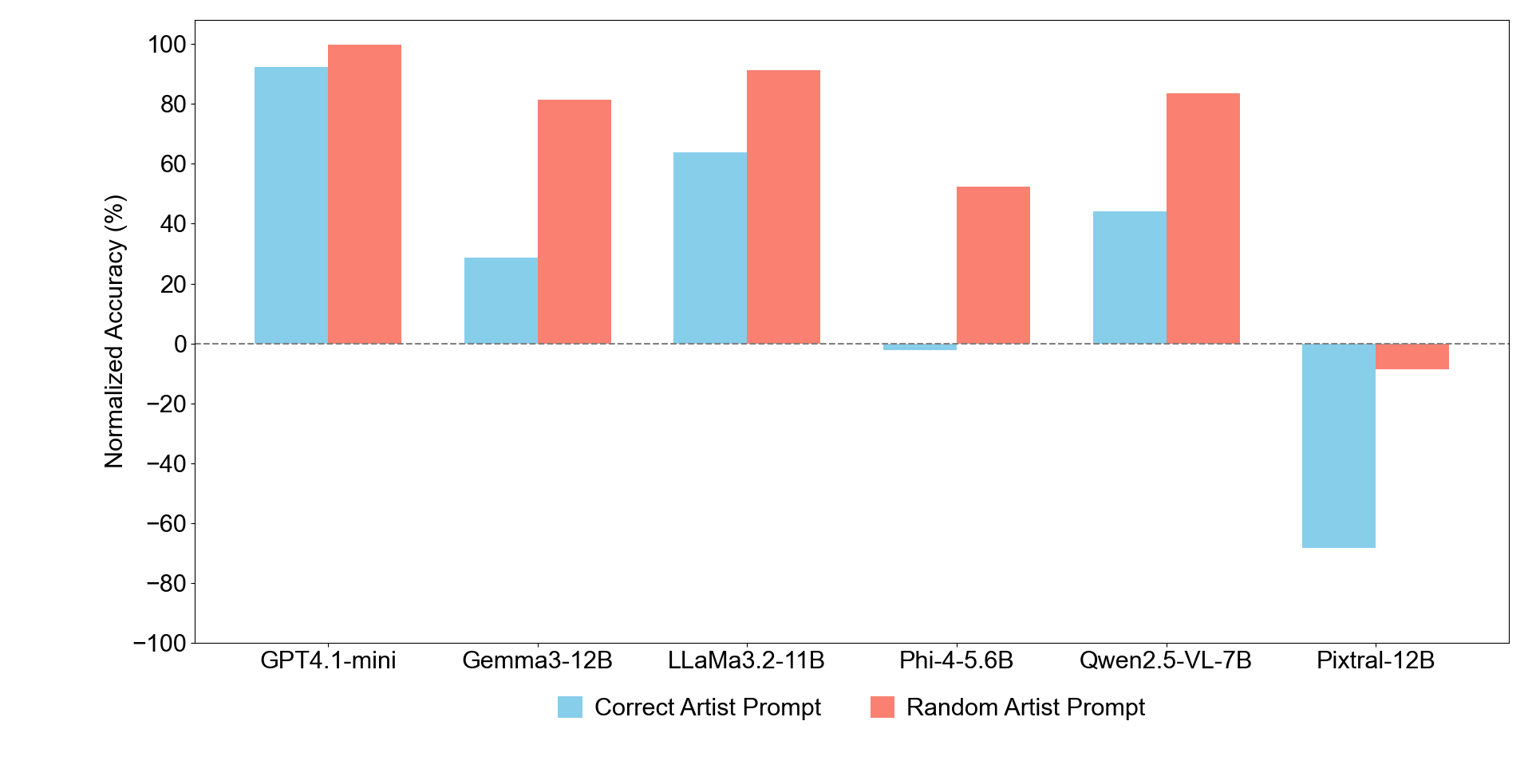}
    \caption{Average $C1$ and $C2$ scores on all painters for the VLMs considered on the images generated with Stable Diffusion}
    \label{fig:C1_C2_SD}
\end{figure}

\begin{figure}[htbp]
    \centering
    \includegraphics[width=0.8\linewidth]{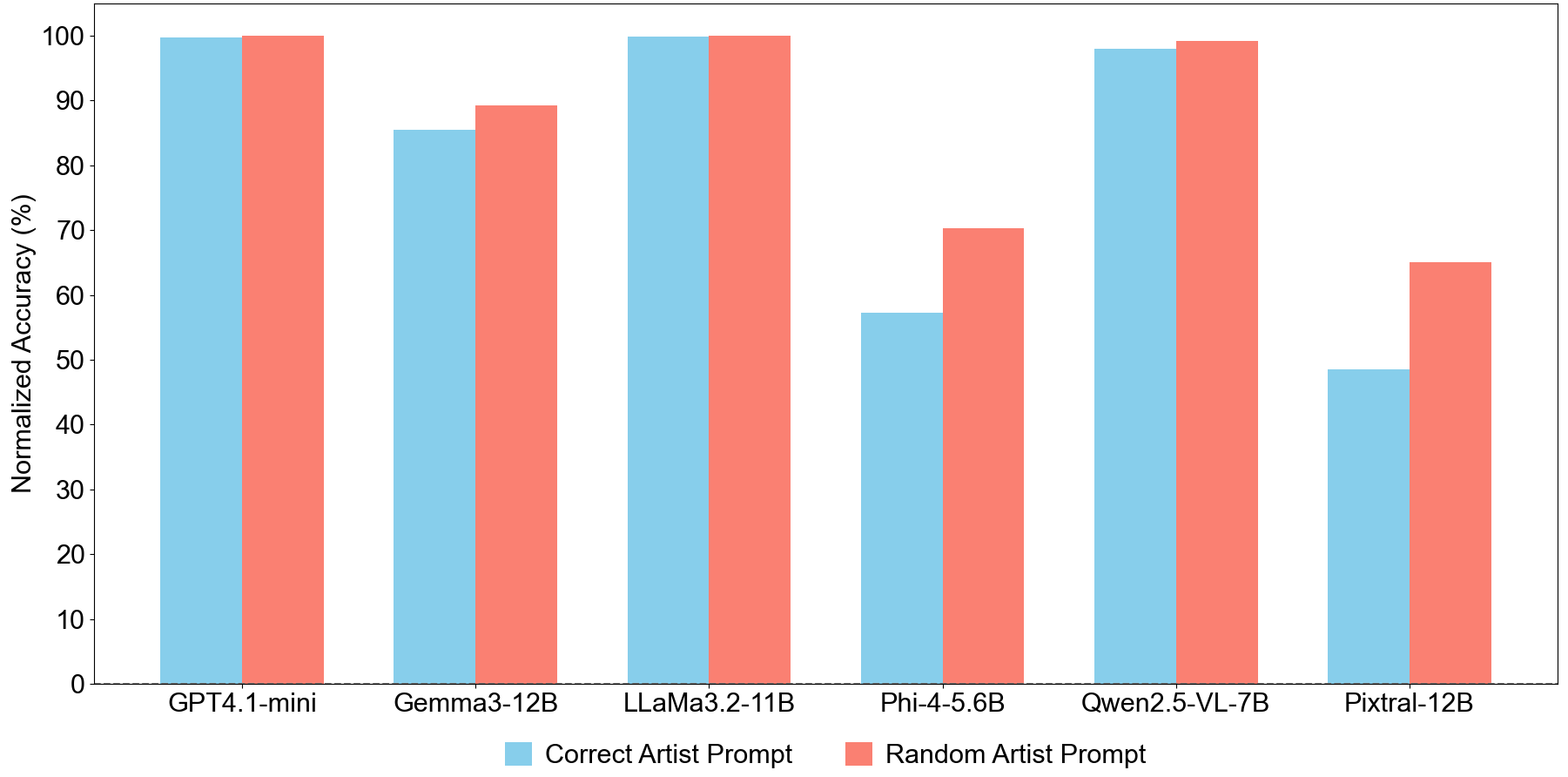}
    \caption{Average $C1$ and $C2$ scores on all painters for the VLMs considered on the images generated with Flux}
    \label{fig:C1_C2_Flux}
\end{figure}

\begin{figure}[htbp]
    \centering
    \includegraphics[width=0.8\linewidth]{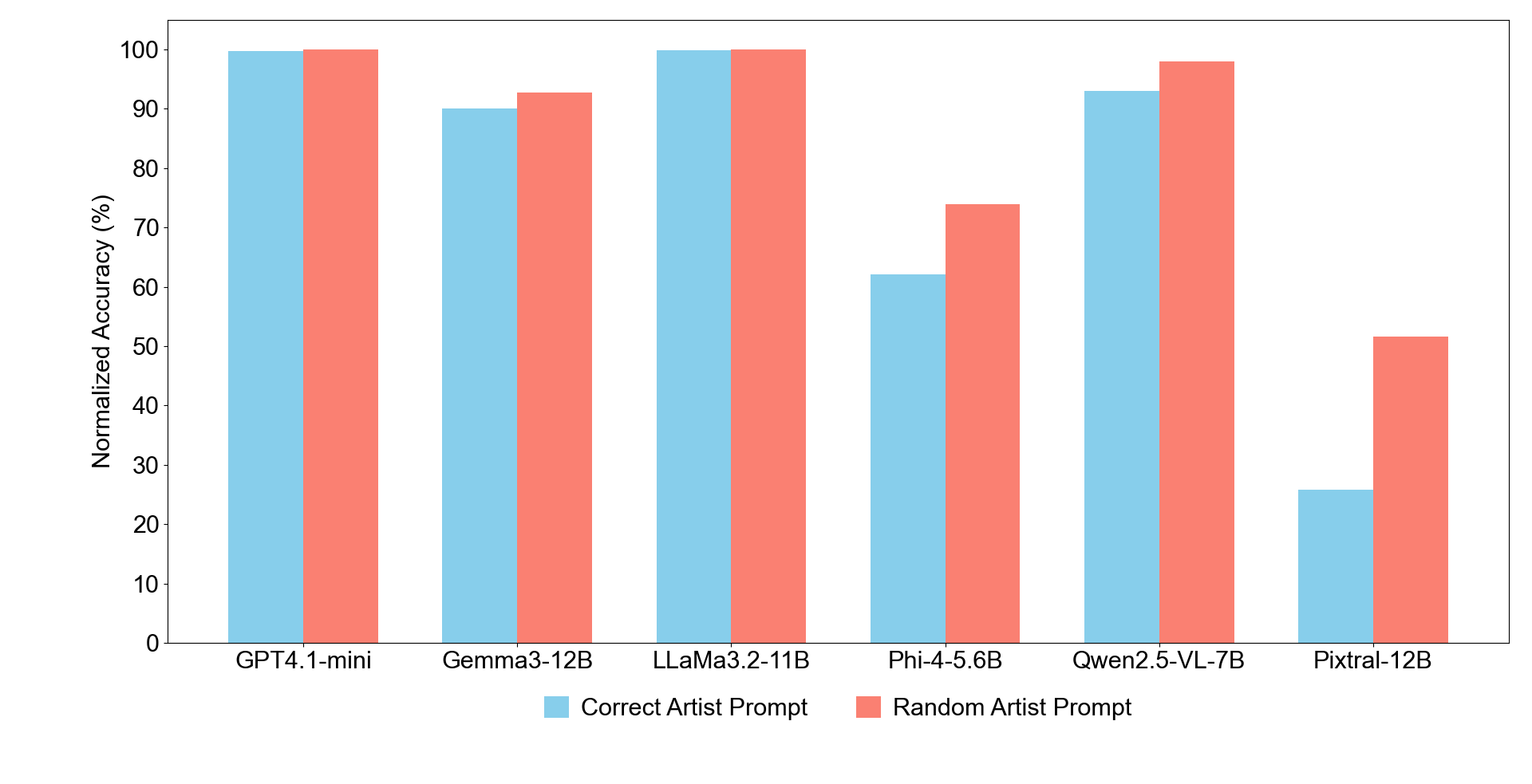}
    \caption{Average $C1$ and $C2$ scores on all painters for the VLMs considered on the images generated with F-Lite}
    \label{fig:C1_C2_F-lite}
\end{figure}

In the case of AI-generated paintings, the correct answer is \textit{no} for both Prompt-1 and Prompt-2. The values of $C1,C2$ are shown per model in Figures \ref{fig:C1_C2_SD}, \ref{fig:C1_C2_Flux}, \ref{fig:C1_C2_F-lite} for Stable Diffusion, Flux, and F-Lite respectively. 

For Stable Diffusion, it can be observed that GPT4.1-mini is the best performing model that is capable of identifying over 95\% of the canvas as not generated by the suggested painter. This is consistent with the behavior observed for real paintings for which GPT4.1-mini was also capable of not attributing a canvas to an incorrect painter. On the other extreme, Pixtral-12B gets the worse scores as it tends to identify the image with the proposed painter. LLaMa3.2-11B also has good results, with the rest of the models obtaining lower values. 

Across models, the results for the painter being imitated by the AI generator are lower than for a random painter. This indicates that models can, to some extent, imitate the style of painters in a way that fools VLMs. This effect is quite large in all models except GPT4.1-mini.

\begin{figure}[htbp]
    \centering
    \includegraphics[width=0.8\linewidth]{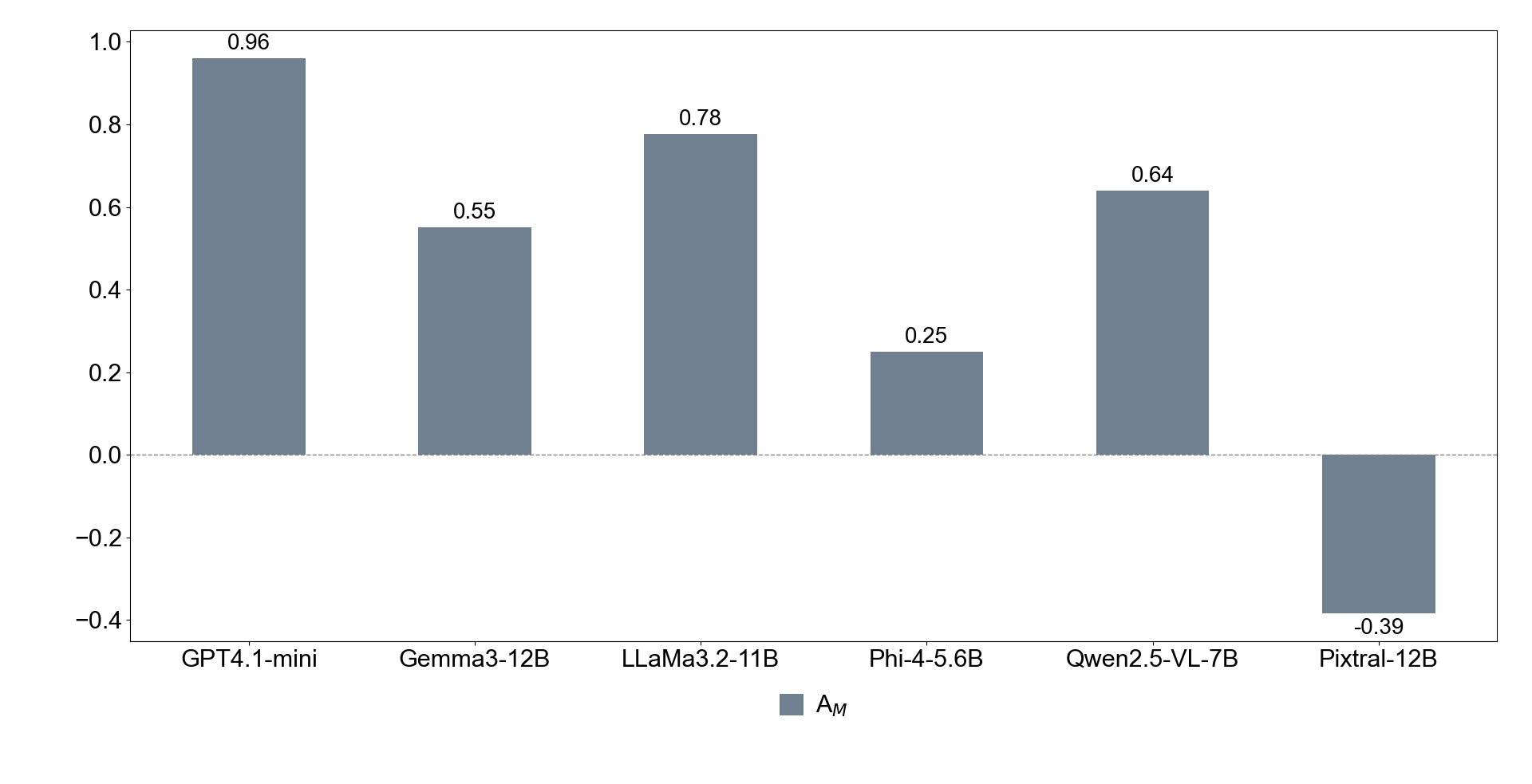}
    \caption{Average $A_M$ scores on all painters for the VLMs considered on the images generated with Stable Diffusion}
    \label{fig:Am_SD}
\end{figure}

The combined results in terms of $A_M$ are shown per model in Figure \ref{fig:Am_SD}. The results show that again GPT4.1-mini is the best performing model and Pixtral-12B the worst. LLaMa3.2-11B also has good performance identifying most of the AI generated images as not being created by a painter.

\begin{figure}[htbp]
    \centering
    \includegraphics[width=0.8\linewidth]{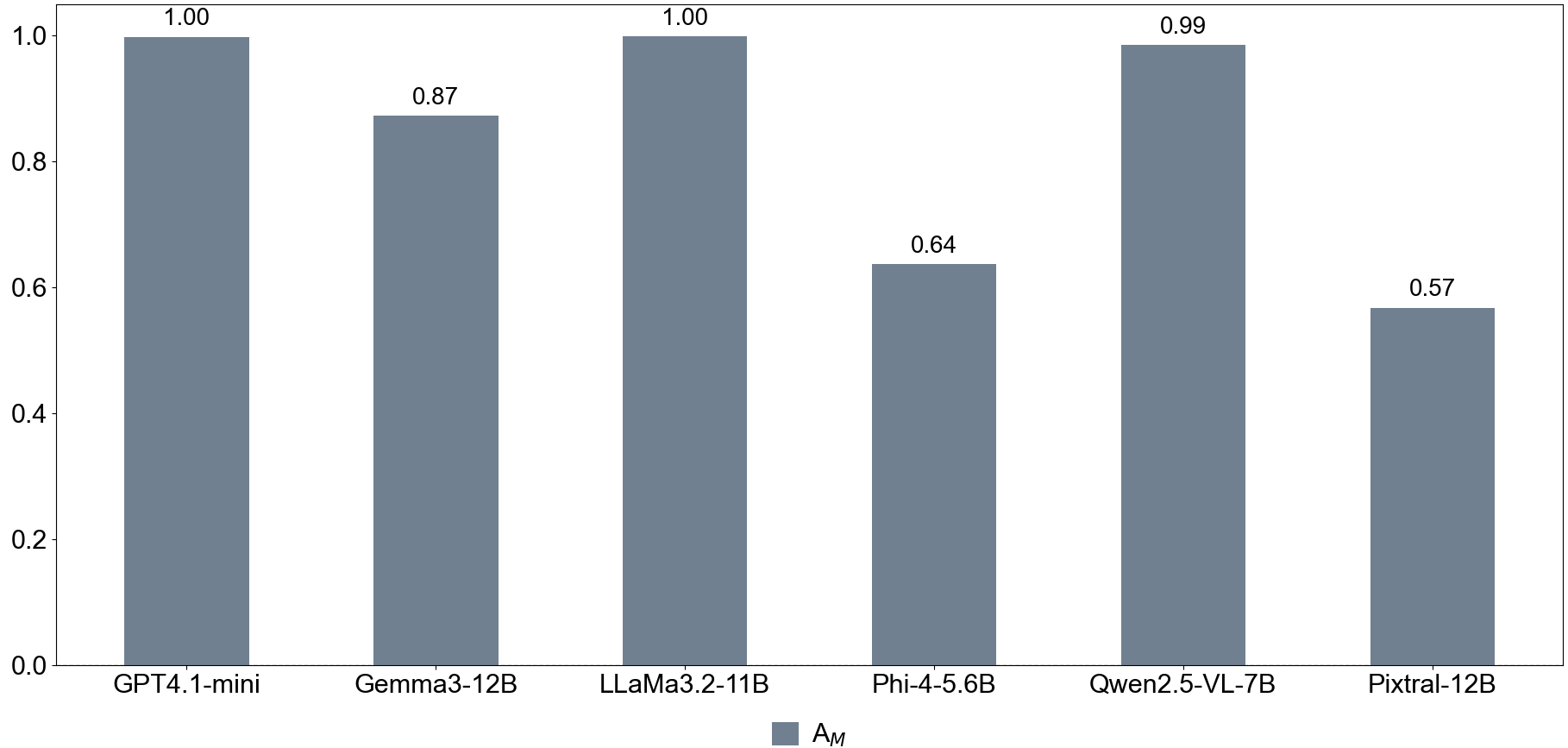}
    \caption{Average $A_M$ scores on all painters for the VLMs considered on the images generated with Flux}
    \label{fig:Am_Flux}
\end{figure}

\begin{figure}[htbp]
    \centering
    \includegraphics[width=0.8\linewidth]{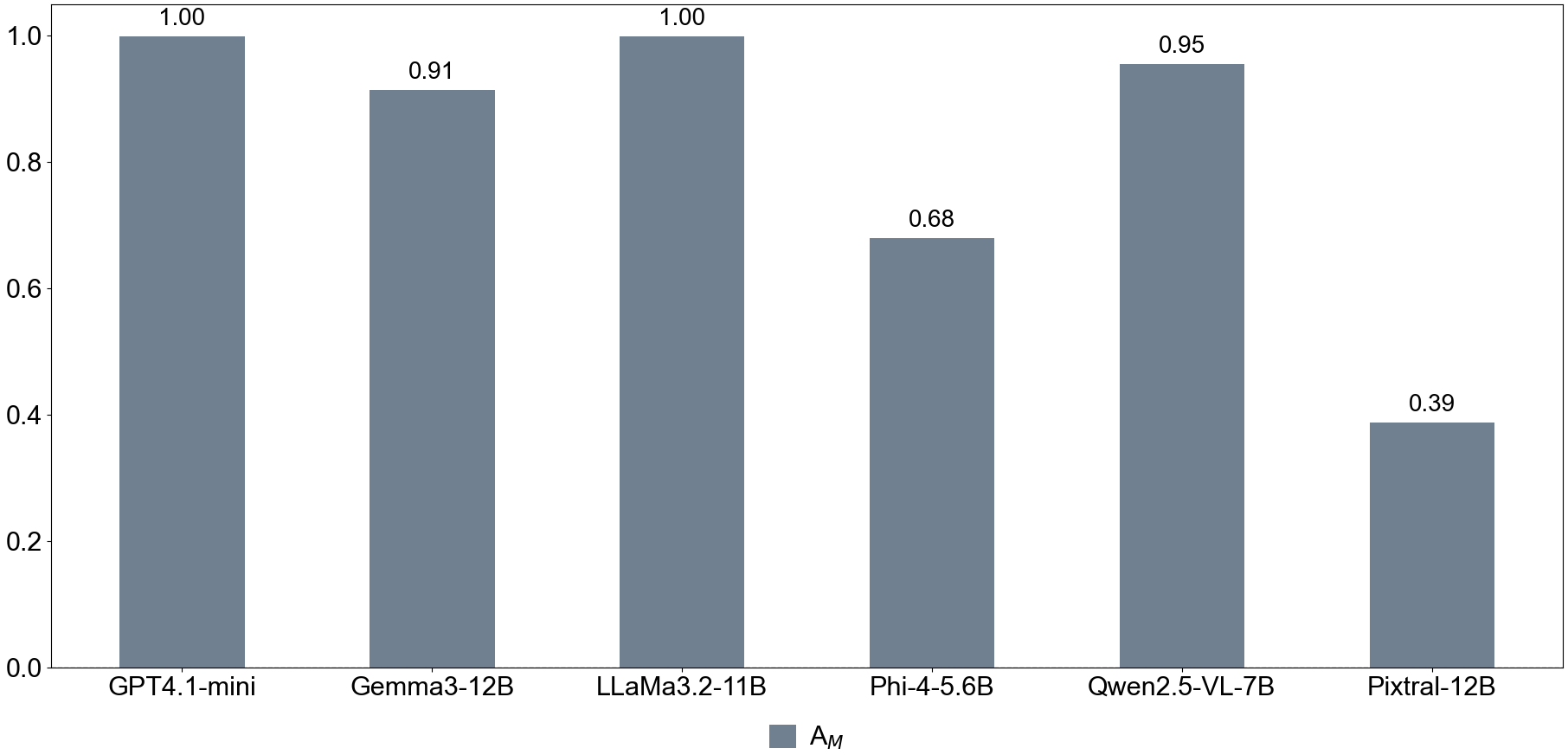}
    \caption{Average $A_M$ scores on all painters for the VLMs considered on the images generated with F-Lite}
    \label{fig:Am_Freepik}
\end{figure}

For Flux and F-Lite, the results are similar and quite different from those of Stable Diffusion. All VLMs can identify the majority of images as not being painted by the proposed artist. In fact, three models, GPT4.1-mini, LLaMa3.2-11B, and Qwen2.5-VL-7B, achieved close to 100\% accuracy. This is clearly seen in Figures \ref{fig:Am_Flux}, \ref{fig:Am_Freepik}. These results suggest that some AI image generators have a style that can be easily recognized as not corresponding to human artists. In this case, the performance gap between the correct and incorrect painter prompts is also smaller, confirming that Flux imitations are easily identified.  

\begin{figure*}[htbp]
    \centering
    \includegraphics[width=\linewidth]{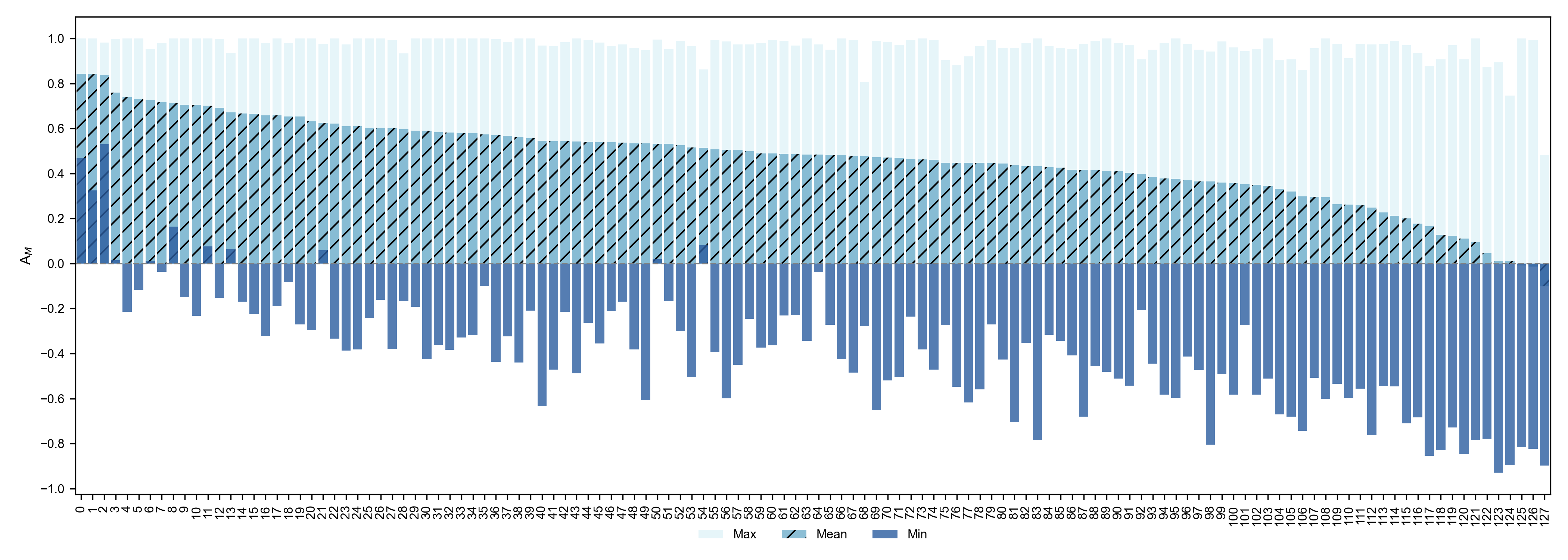}
    \caption{Painters ordered by average $A_M$ scores for the VLMs considered on the images generated with Stable Diffusion. The maximum and minimum $A_M$ scores across the VLMs are also shown.}
    \label{fig:artists_AM_SD}
\end{figure*}

\begin{figure*}[htbp]
    \centering
    \includegraphics[width=\linewidth]{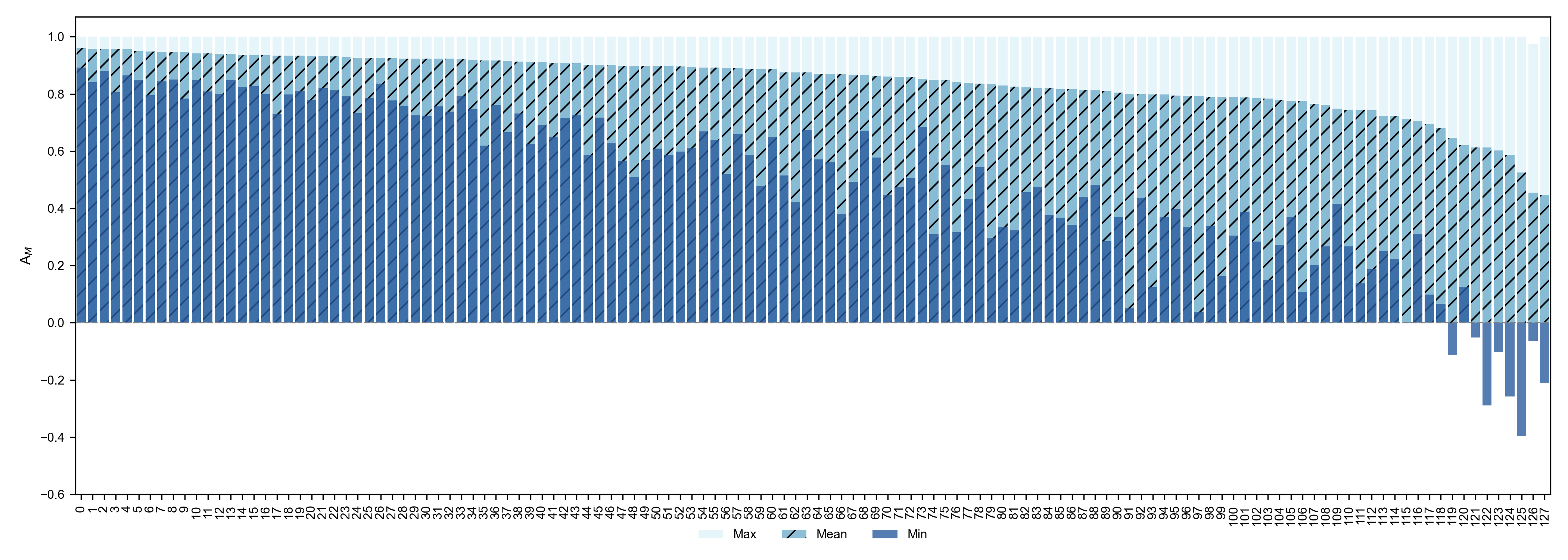}
    \caption{Painters ordered by average $A_M$ scores for the VLMs considered on the images generated with Flux. The maximum and minimum $A_M$ scores across the VLMs are also shown.}
    \label{fig:artists_AM_Flux}
\end{figure*}

\begin{figure*}[htbp]
    \centering
    \includegraphics[width=\linewidth]{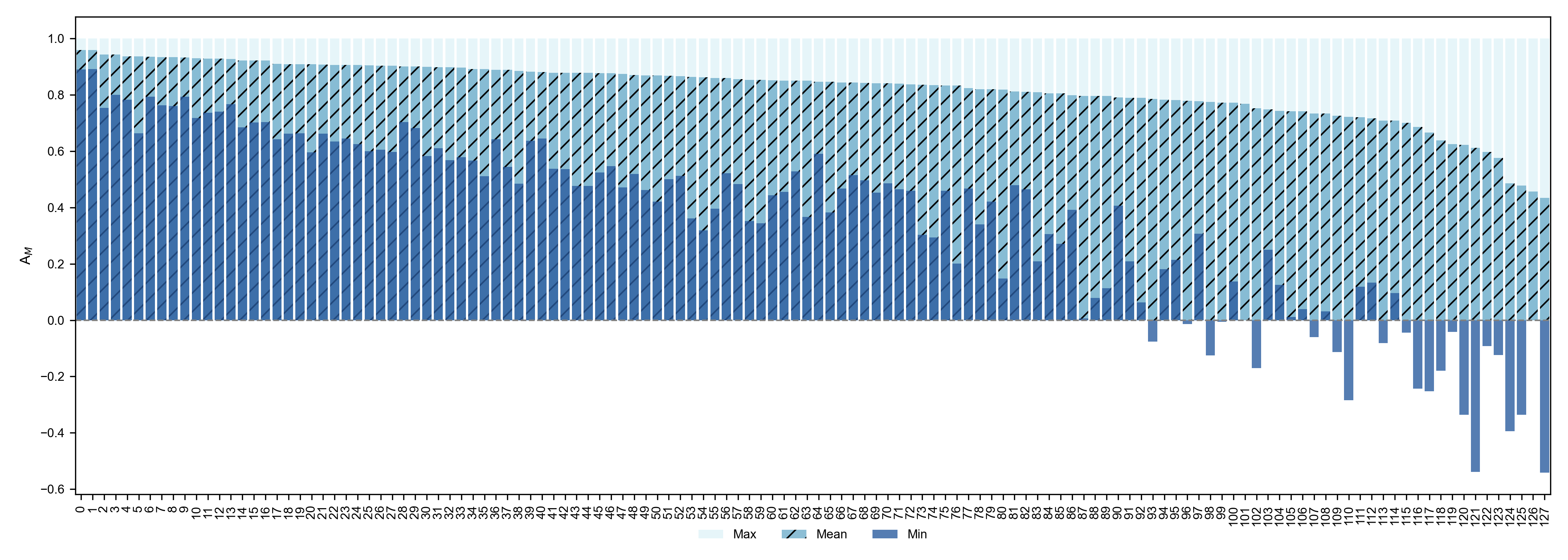}
    \caption{Painters ordered by average $A_M$ scores for the VLMs considered on the images generated with F-Lite. The maximum and minimum $A_M$ scores across the VLMs are also shown.}
    \label{fig:artists_AM_Freepik}
\end{figure*}

The results per painter are shown in Figures \ref{fig:artists_AM_SD},\ref{fig:artists_AM_Flux},\ref{fig:artists_AM_Freepik} with the correspondence of numbers with artists in Tables \ref{tab:artist_list_SD}, \ref{tab:artist_list_Flux}, \ref{tab:artist_list_F-Lite}. 

For Stable Diffusion, M.C. Escher has the best performance. Interestingly, this artist had the worst performance for real paintings. The worst results for Stable Diffusion are for Gustave Loiseau and like with real paintings there is a large difference among the best and worst painters. In this case, there are several popular painters, such as Michelangelo, Leonardo Da Vinci, Salvador Dalí, El Greco, or Andy Warhol among the top 10 which suggest that there can be a relationship between popularity and performance. Another interesting observation is that the spread between the performance of the best and worst models is much larger than for real paintings. This is in part due to the poor performance of Pixtral-12B, which has a negative normalized accuracy.

For Flux, the results are more consistent with smaller differences between the painters and also between the best and worst models. The best performing artist is Henri De Toulouse Lautrec and the worst Maxime Maufra who still is above 40\% average normalized accuracy. For the top performing artists, the worst model is above 80\% average normalized accuracy, so all models can identify the images as not being paintings of the suggested artist. 

For F-Lite, the overall results are similar to those of Flux with smaller differences between the painters and also between the best and worst models than in Stable Diffusion. The best performing artist is Leonardo Da Vinci with Michelangelo, Salvador Dalí, Rembrandt, or Andy Warhol in the top 10. Once again, this may indicate a correlation between artist popularity and performance. Interestingly, M.C. Escher is the third best performing artist. The artist with the worst performance is Antoine Blanchard, and the bottom 10 artists are not among the most popular.

\begin{table*}[htbp]
  \centering
  \tiny
  \begin{tabular}{rlrlrlrl}
    \toprule
    No. & Artist & No. & Artist & No. & Artist & No. & Artist \\
    \midrule
    1 & M.C. Escher & 33 & Martiros Saryan & 65 & Gene Davis & 97 & Henri De Toulouse Lautrec \\
    2 & Michelangelo & 34 & Mstislav Dobuzhinsky & 66 & Hans Holbein The Younger & 98 & Kuzma Petrov Vodkin \\
    3 & Leonardo Da Vinci & 35 & Vasily Vereshchagin & 67 & Jacob Jordaens & 99 & Claude Monet \\
    4 & Salvador Dalí & 36 & Bartolome Esteban Murillo & 68 & James Mcneill Whistler & 100 & Mary Cassatt \\
    5 & El Greco & 37 & Pyotr Konchalovsky & 69 & Fernando Botero & 101 & Aleksey Savrasov \\
    6 & Mikhail Vrubel & 38 & Dante Gabriel Rossetti & 70 & Edouard Manet & 102 & Lucas Cranach The Elder \\
    7 & Andy Warhol & 39 & Jacek Malczewski & 71 & Edgar Degas & 103 & Ilya Repin \\
    8 & Albrecht Durer & 40 & Hieronymus Bosch & 72 & Karl Bryullov & 104 & Ernst Ludwig Kirchner \\
    9 & Ivan Bilibin & 41 & Arkhip Kuindzhi & 73 & Nicholas Roerich & 105 & Isaac Levitan \\
    10 & Jan Matejko & 42 & Frans Hals & 74 & Vasily Perov & 106 & Vasily Polenov \\
    11 & Peter Paul Rubens & 43 & Gustave Dore & 75 & Zinaida Serebriakova & 107 & Paul Cezanne \\
    12 & Sam Francis & 44 & Vasily Surikov & 76 & Edward Burne Jones & 108 & Henri Matisse \\
    13 & Rembrandt & 45 & Anthony Van Dyck & 77 & Canaletto & 109 & Ilya Mashkov \\
    14 & Katsushika Hokusai & 46 & Ivan Shishkin & 78 & Thomas Eakins & 110 & Raoul Dufy \\
    15 & Gustave Moreau & 47 & Georges Braque & 79 & Antoine Blanchard & 111 & Ferdinand Hodler \\
    16 & Tintoretto & 48 & Marc Chagall & 80 & Boris Kustodiev & 112 & Berthe Morisot \\
    17 & Francisco Goya & 49 & Valentin Serov & 81 & Theo Van Rysselberghe & 113 & Camille Corot \\
    18 & Odilon Redon & 50 & Lucian Freud & 82 & Joaquã­N Sorolla & 114 & Ivan Aivazovsky \\
    19 & Raphael & 51 & Utagawa Kuniyoshi & 83 & Juan Gris & 115 & Nikolay Bogdanov Belsky \\
    20 & Mikalojus Ciurlionis & 52 & Fra Angelico & 84 & Paul Gauguin & 116 & Maurice Prendergast \\
    21 & Viktor Vasnetsov & 53 & Orest Kiprensky & 85 & Koloman Moser & 117 & William Merritt Chase \\
    22 & Vincent Van Gogh & 54 & Egon Schiele & 86 & Felix Vallotton & 118 & Konstantin Korovin \\
    23 & Titian & 55 & Raphael Kirchner & 87 & Ivan Kramskoy & 119 & Gustave Caillebotte \\
    24 & Eugene Delacroix & 56 & Joseph Wright & 88 & Childe Hassam & 120 & Pierre Auguste Renoir \\
    25 & Giovanni Boldini & 57 & Gustave Courbet & 89 & Henri Fantin Latour & 121 & John Henry Twachtman \\
    26 & William Turner & 58 & John Singer Sargent & 90 & Konstantin Makovsky & 122 & Henri Martin \\
    27 & Gustav Klimt & 59 & Hans Memling & 91 & Amedeo Modigliani & 123 & Camille Pissarro \\
    28 & Thomas Gainsborough & 60 & Edvard Munch & 92 & Edouard Cortes & 124 & Alfred Sisley \\
    29 & Pablo Picasso & 61 & James Tissot & 93 & Pietro Perugino & 125 & Eugene Boudin \\
    30 & Aubrey Beardsley & 62 & Joshua Reynolds & 94 & Georges Seurat & 126 & Pierre Bonnard \\
    31 & Niko Pirosmani & 63 & Fernand Leger & 95 & Konstantin Somov & 127 & Maxime Maufra \\
    32 & Paolo Veronese & 64 & Sir Lawrence Alma Tadema & 96 & David Burliuk & 128 & Gustave Loiseau \\
    \bottomrule
  \end{tabular}
  \caption{List of 128 Artists Sorted by Arithmetic Mean Normalized Accuracy of Stable Diffusion-Generated paintings}
  \label{tab:artist_list_SD}
\end{table*}

\begin{table*}[htbp]
  \centering
  \tiny
  \begin{tabular}{rlrlrlrl}
    \toprule
    No. & Artist & No. & Artist & No. & Artist & No. & Artist \\
    \midrule
    1 & Henri De Toulouse Lautrec & 33 & Karl Bryullov & 65 & Paul Gauguin & 97 & Vasily Polenov \\
    2 & Dante Gabriel Rossetti & 34 & Berthe Morisot & 66 & Camille Pissarro & 98 & Ilya Repin \\
    3 & Mary Cassatt & 35 & Vasily Surikov & 67 & Tintoretto & 99 & Raphael Kirchner \\
    4 & Michelangelo & 36 & Gustave Dore & 68 & Boris Kustodiev & 100 & Claude Monet \\
    5 & Edgar Degas & 37 & Henri Matisse & 69 & Ferdinand Hodler & 101 & Fernand Leger \\
    6 & Giovanni Boldini & 38 & Viktor Vasnetsov & 70 & Konstantin Somov & 102 & Ivan Shishkin \\
    7 & Sir Lawrence Alma Tadema & 39 & Odilon Redon & 71 & Gustav Klimt & 103 & Nicholas Roerich \\
    8 & Ivan Kramskoy & 40 & Albrecht Durer & 72 & Raphael & 104 & Jacob Jordaens \\
    9 & Lucian Freud & 41 & Salvador Dalí & 73 & Jacek Malczewski & 105 & Hans Memling \\
    10 & Jan Matejko & 42 & Rembrandt & 74 & Mstislav Dobuzhinsky & 106 & Raoul Dufy \\
    11 & Joaquã­N Sorolla & 43 & James Mcneill Whistler & 75 & Pierre Bonnard & 107 & Camille Corot \\
    12 & Amedeo Modigliani & 44 & Katsushika Hokusai & 76 & Pyotr Konchalovsky & 108 & Lucas Cranach The Elder \\
    13 & Edouard Manet & 45 & Hans Holbein The Younger & 77 & Arkhip Kuindzhi & 109 & Isaac Levitan \\
    14 & Leonardo Da Vinci & 46 & Niko Pirosmani & 78 & Georges Seurat & 110 & William Merritt Chase \\
    15 & John Singer Sargent & 47 & M.C. Escher & 79 & Fernando Botero & 111 & Edouard Cortes \\
    16 & Edvard Munch & 48 & Anthony Van Dyck & 80 & Gustave Caillebotte & 112 & Konstantin Korovin \\
    17 & Mikhail Vrubel & 49 & Fra Angelico & 81 & Hieronymus Bosch & 113 & Mikalojus Ciurlionis \\
    18 & El Greco & 50 & Zinaida Serebriakova & 82 & Thomas Gainsborough & 114 & Henri Martin \\
    19 & Vasily Perov & 51 & Thomas Eakins & 83 & Theo Van Rysselberghe & 115 & Ivan Aivazovsky \\
    20 & Pablo Picasso & 52 & Peter Paul Rubens & 84 & Maurice Prendergast & 116 & Alfred Sisley \\
    21 & Francisco Goya & 53 & Titian & 85 & Utagawa Kuniyoshi & 117 & Juan Gris \\
    22 & Marc Chagall & 54 & Paul Cezanne & 86 & Ernst Ludwig Kirchner & 118 & Sam Francis \\
    23 & Vasily Vereshchagin & 55 & Kuzma Petrov Vodkin & 87 & Henri Fantin Latour & 119 & Aleksey Savrasov \\
    24 & Vincent Van Gogh & 56 & Georges Braque & 88 & Martiros Saryan & 120 & Antoine Blanchard \\
    25 & Eugene Delacroix & 57 & Pietro Perugino & 89 & William Turner & 121 & Nikolay Bogdanov Belsky \\
    26 & Edward Burne Jones & 58 & Pierre Auguste Renoir & 90 & Canaletto & 122 & Ilya Mashkov \\
    27 & Egon Schiele & 59 & Childe Hassam & 91 & Bartolome Esteban Murillo & 123 & John Henry Twachtman \\
    28 & Orest Kiprensky & 60 & Frans Hals & 92 & Aubrey Beardsley & 124 & David Burliuk \\
    29 & Valentin Serov & 61 & Gustave Courbet & 93 & Koloman Moser & 125 & Gustave Loiseau \\
    30 & Ivan Bilibin & 62 & Paolo Veronese & 94 & Joshua Reynolds & 126 & Eugene Boudin \\
    31 & Gustave Moreau & 63 & Andy Warhol & 95 & Joseph Wright & 127 & Gene Davis \\
    32 & James Tissot & 64 & Felix Vallotton & 96 & Konstantin Makovsky & 128 & Maxime Maufra \\
    \bottomrule
  \end{tabular}
  \caption{List of 128 Artists Sorted by Arithmetic Mean Normalized Accuracy of Flux-Generated paintings}
  \label{tab:artist_list_Flux}
\end{table*}

\begin{table*}[htbp]
  \centering
  \tiny
  \begin{tabular}{rlrlrlrl}
    \toprule
    No. & Artist & No. & Artist & No. & Artist & No. & Artist \\
    \midrule
    1 & Leonardo Da Vinci & 33 & Mikhail Vrubel & 65 & Henri Fantin Latour & 97 & Paul Gauguin \\
    2 & Michelangelo & 34 & Titian & 66 & Berthe Morisot & 98 & Sir Lawrence Alma Tadema \\
    3 & M.C. Escher & 35 & Joaquã­N Sorolla & 67 & Utagawa Kuniyoshi & 99 & Claude Monet \\
    4 & Salvador Dalí & 36 & Dante Gabriel Rossetti & 68 & Ferdinand Hodler & 100 & Aubrey Beardsley \\
    5 & Rembrandt & 37 & Edward Burne Jones & 69 & Ivan Bilibin & 101 & Isaac Levitan \\
    6 & Andy Warhol & 38 & Peter Paul Rubens & 70 & Canaletto & 102 & Jacob Jordaens \\
    7 & Edvard Munch & 39 & Fra Angelico & 71 & Bartolome Esteban Murillo & 103 & Gustave Caillebotte \\
    8 & Raphael & 40 & Katsushika Hokusai & 72 & Zinaida Serebriakova & 104 & Juan Gris \\
    9 & Eugene Delacroix & 41 & William Turner & 73 & Raphael Kirchner & 105 & Ivan Shishkin \\
    10 & Amedeo Modigliani & 42 & Giovanni Boldini & 74 & Georges Seurat & 106 & Theo Van Rysselberghe \\
    11 & Albrecht Durer & 43 & Vasily Vereshchagin & 75 & Ernst Ludwig Kirchner & 107 & Aleksey Savrasov \\
    12 & Marc Chagall & 44 & Paolo Veronese & 76 & Joseph Wright & 108 & Joshua Reynolds \\
    13 & Henri De Toulouse Lautrec & 45 & Tintoretto & 77 & Lucian Freud & 109 & Vasily Polenov \\
    14 & El Greco & 46 & James Tissot & 78 & Gustav Klimt & 110 & Pierre Bonnard \\
    15 & Edgar Degas & 47 & Vasily Surikov & 79 & Hieronymus Bosch & 111 & Alfred Sisley \\
    16 & Pablo Picasso & 48 & Valentin Serov & 80 & Ivan Aivazovsky & 112 & Hans Memling \\
    17 & Mary Cassatt & 49 & Georges Braque & 81 & Childe Hassam & 113 & William Merritt Chase \\
    18 & James Mcneill Whistler & 50 & Henri Matisse & 82 & Fernand Leger & 114 & Ilya Repin \\
    19 & Vincent Van Gogh & 51 & Hans Holbein The Younger & 83 & Boris Kustodiev & 115 & Maurice Prendergast \\
    20 & Egon Schiele & 52 & Viktor Vasnetsov & 84 & Jacek Malczewski & 116 & Lucas Cranach The Elder \\
    21 & Thomas Eakins & 53 & Odilon Redon & 85 & Koloman Moser & 117 & Konstantin Korovin \\
    22 & Sam Francis & 54 & Gustave Moreau & 86 & Mikalojus Ciurlionis & 118 & Eugene Boudin \\
    23 & Francisco Goya & 55 & Frans Hals & 87 & Nicholas Roerich & 119 & Konstantin Makovsky \\
    24 & Felix Vallotton & 56 & Gustave Courbet & 88 & Gustave Dore & 120 & John Henry Twachtman \\
    25 & John Singer Sargent & 57 & Niko Pirosmani & 89 & Camille Corot & 121 & Henri Martin \\
    26 & Karl Bryullov & 58 & Kuzma Petrov Vodkin & 90 & Konstantin Somov & 122 & Gustave Loiseau \\
    27 & Edouard Manet & 59 & Pietro Perugino & 91 & Gene Davis & 123 & David Burliuk \\
    28 & Thomas Gainsborough & 60 & Pyotr Konchalovsky & 92 & Martiros Saryan & 124 & Ilya Mashkov \\
    29 & Orest Kiprensky & 61 & Fernando Botero & 93 & Camille Pissarro & 125 & Maxime Maufra \\
    30 & Jan Matejko & 62 & Pierre Auguste Renoir & 94 & Paul Cezanne & 126 & Edouard Cortes \\
    31 & Anthony Van Dyck & 63 & Vasily Perov & 95 & Raoul Dufy & 127 & Nikolay Bogdanov Belsky \\
    32 & Ivan Kramskoy & 64 & Arkhip Kuindzhi & 96 & Mstislav Dobuzhinsky & 128 & Antoine Blanchard \\
    \bottomrule
  \end{tabular}
  \caption{List of 128 Artists Sorted by Arithmetic Mean Normalized Accuracy of F-Lite-Generated paintings}
  \label{tab:artist_list_F-Lite}
\end{table*}

\begin{figure*}[htbp]
    \centering
    \includegraphics[width=.4\linewidth]{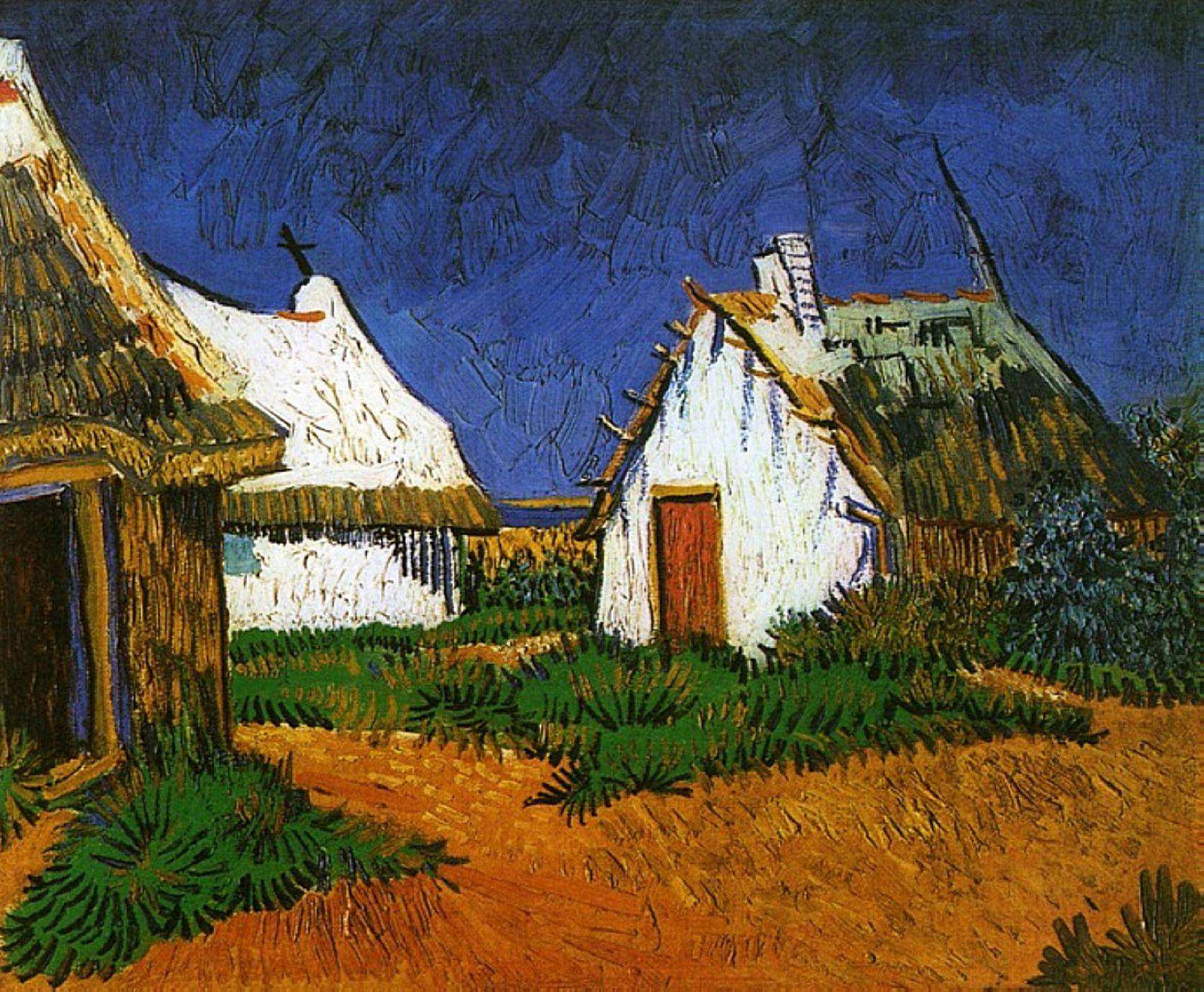}
    \includegraphics[width=.35\linewidth]{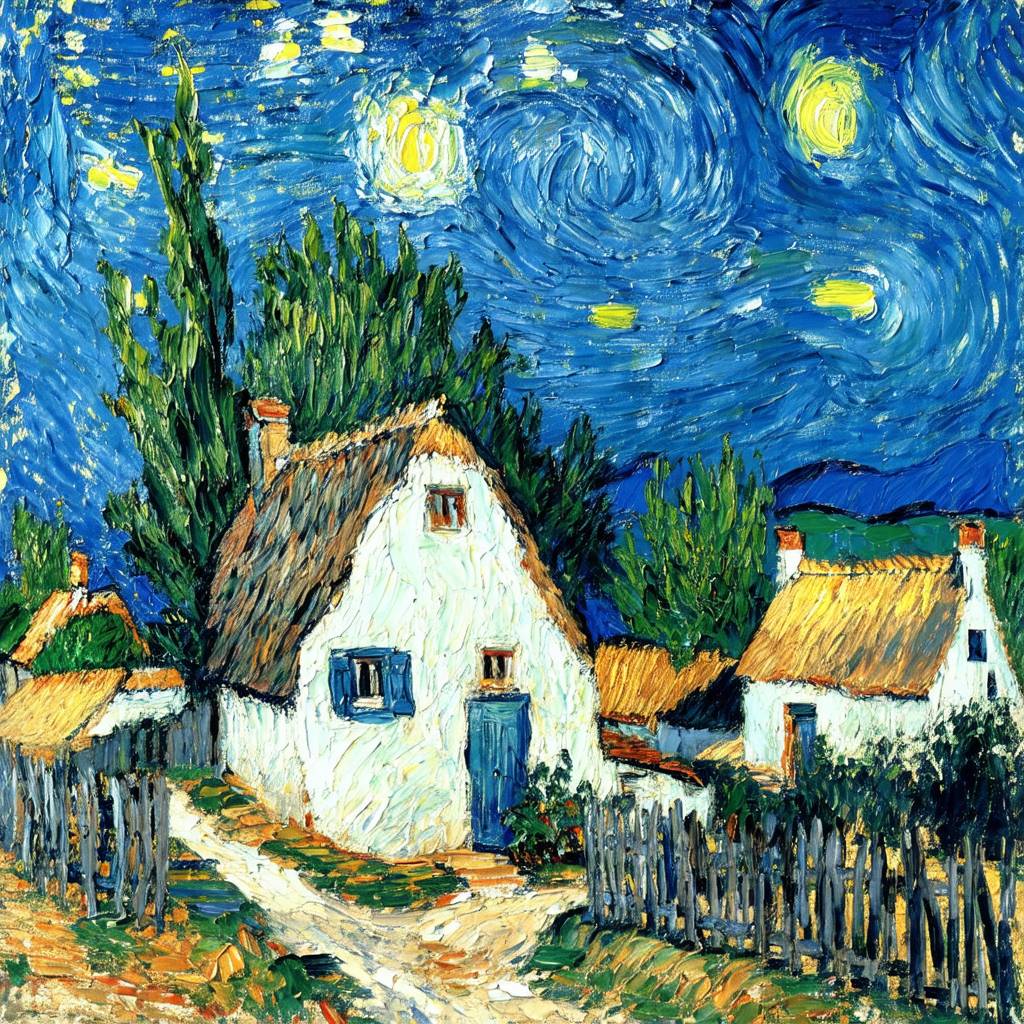}
    \includegraphics[width=.38\linewidth]{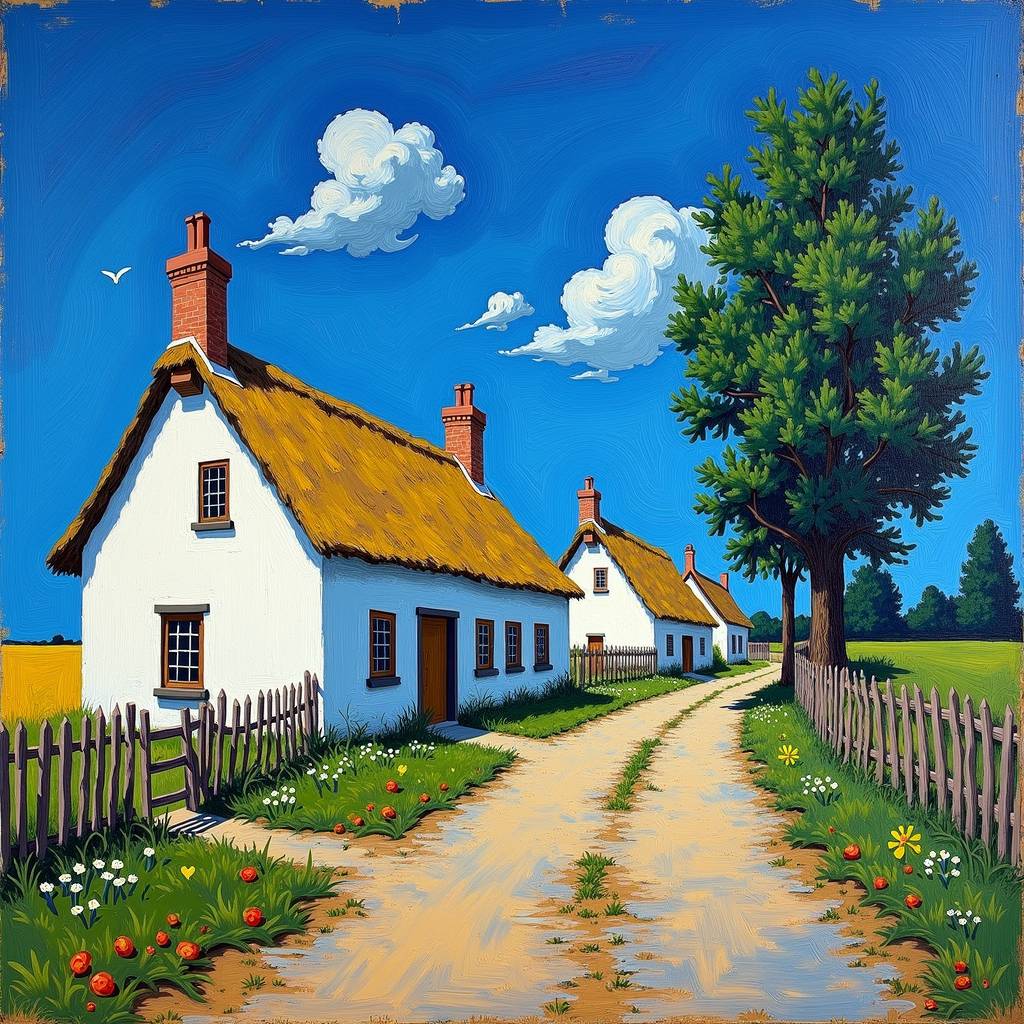}
    \includegraphics[width=.38\linewidth]{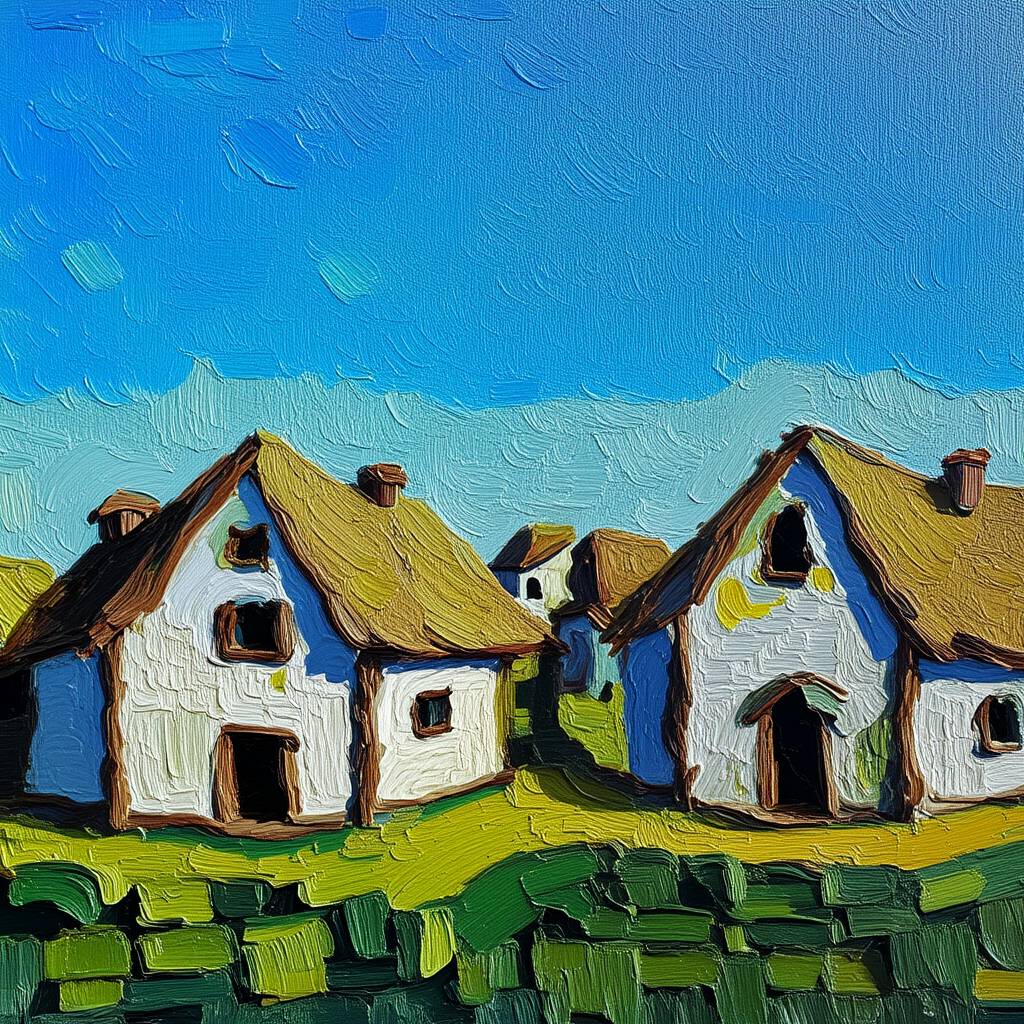}
    \caption{A painting by Van Gogh (top-left) and the images generated by Stable Diffusion (top-right), Flux (bottom-left) and F-Lite (bottom-right) .}
    \label{fig:VanGogh}
\end{figure*}

\begin{figure}[h]
    \centering
    \includegraphics[width=0.4\textwidth]{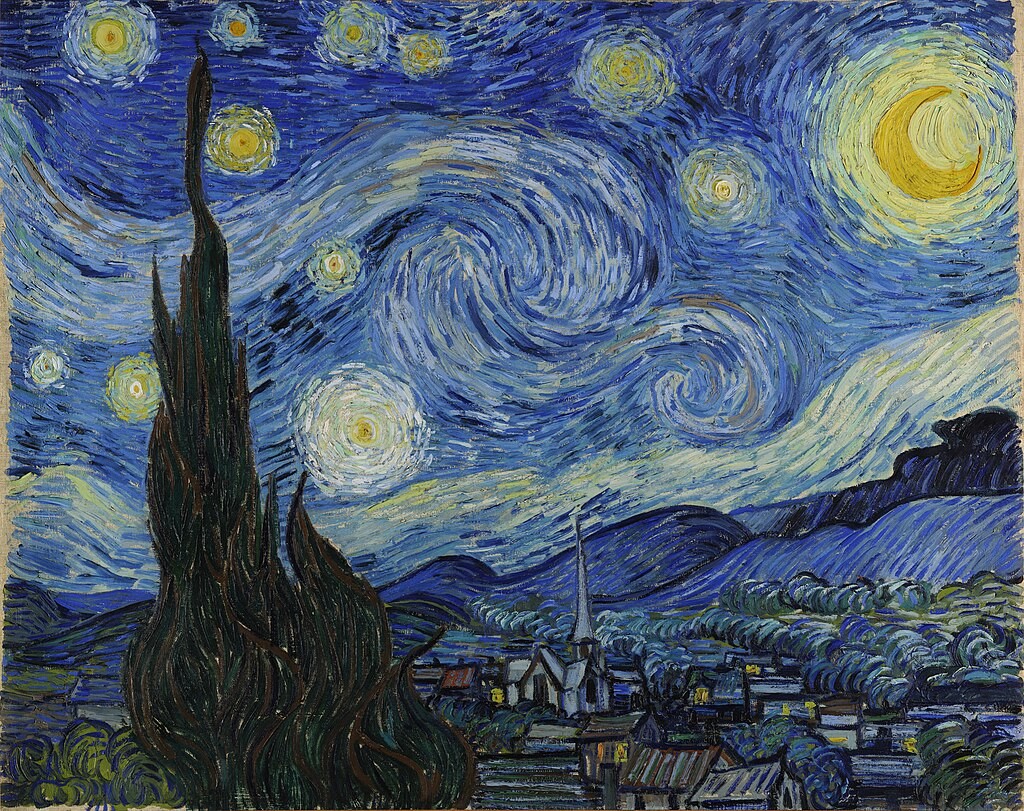}
    \caption{Vincent Van Gogh. 1889. Starry Night. Museum of Moder Art. New York (source \url{https://artsandculture.google.com/asset/bgEuwDxel93-Pg}).}
    \label{fig:starry}
\end{figure}

\begin{figure}[h]
    \centering
    \includegraphics[width=0.4\textwidth]{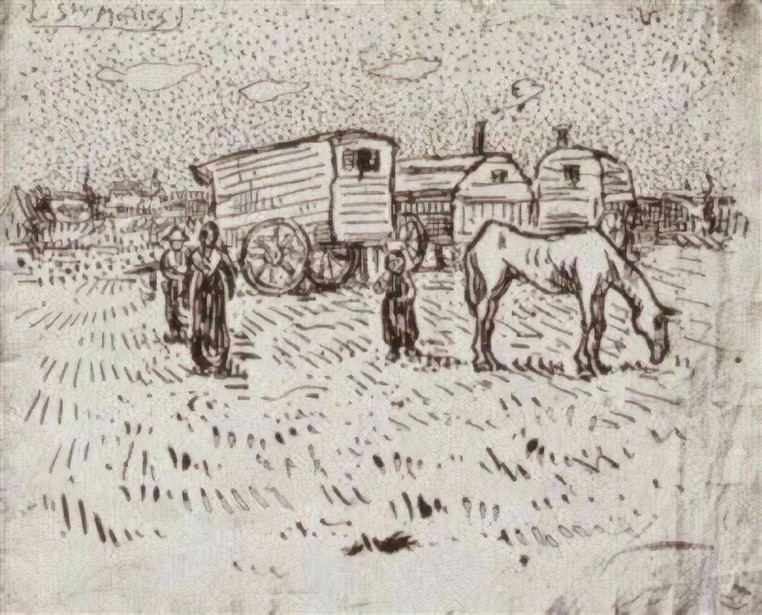}
    \includegraphics[width=0.35\textwidth]{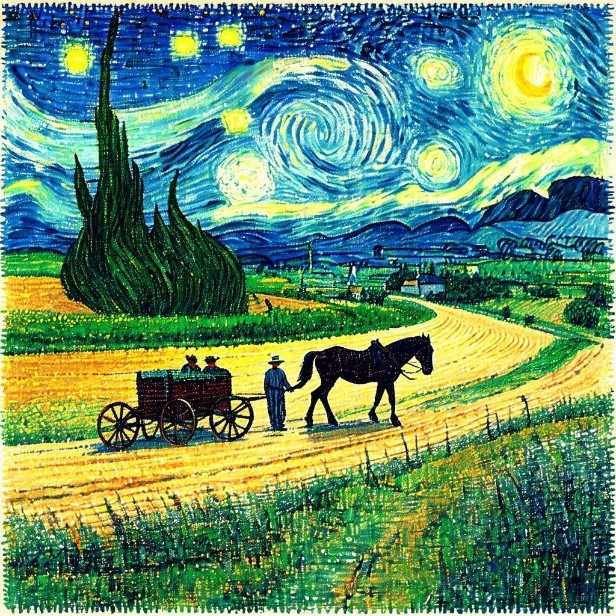}
    \includegraphics[width=0.4\textwidth]{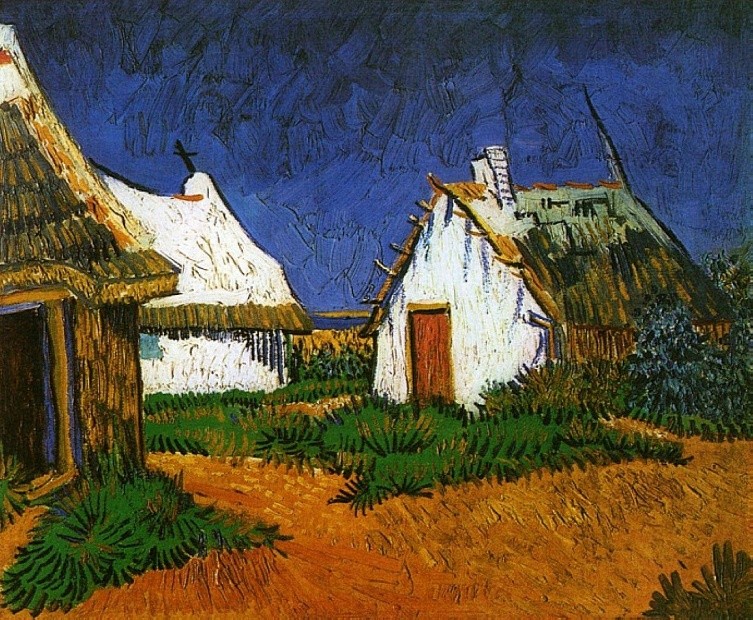}
    \includegraphics[width=0.35\textwidth]{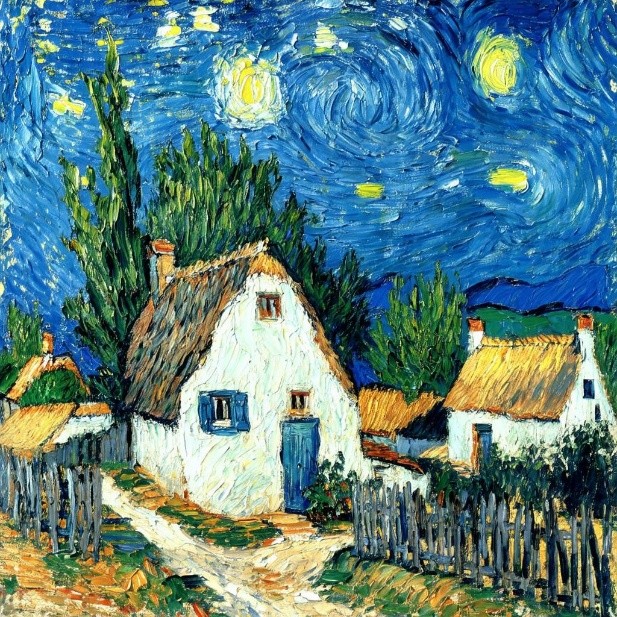}
    \includegraphics[width=0.4\textwidth]{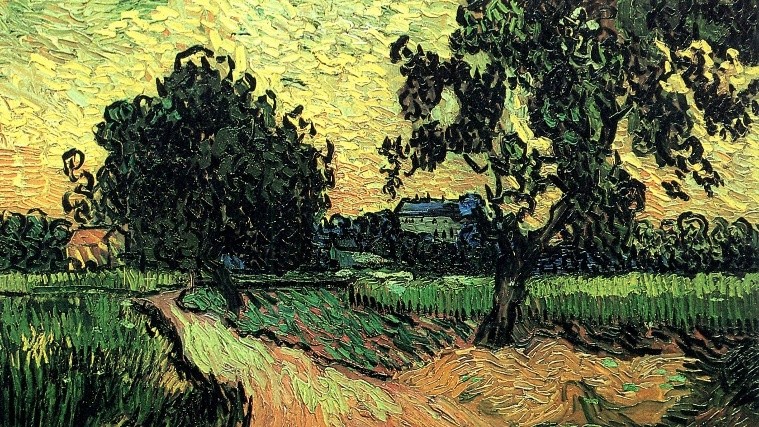}
    \includegraphics[width=0.35\textwidth]{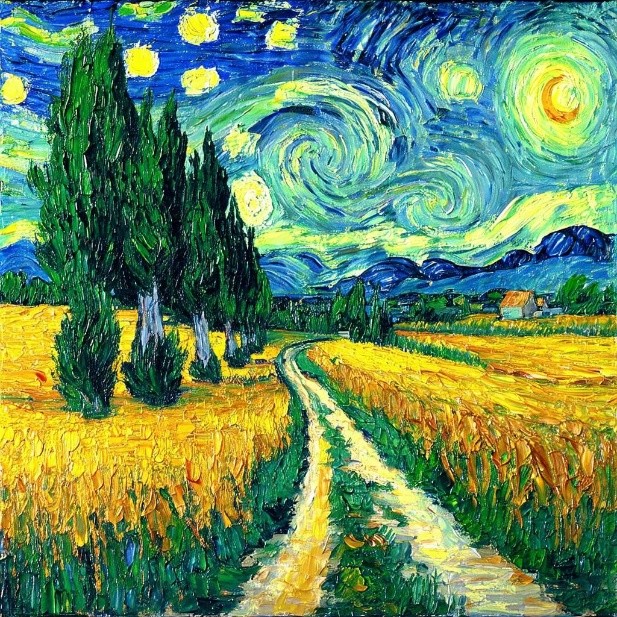}
    \caption{On the left are original works by Vincent van Gogh; on the right, parallel recreations by Stable Diffusion, which incorporate the characteristic swirling of Starry Night despite its absence in the original images. (Items 49, 51 and 65 in  \url{https://ama2210.github.io/WikiArt_VLM_Web/}.}
    \label{fig:starrySD}
\end{figure}

To illustrate the differences between real and AI-generated images, an example is shown in Figure\ref{fig:VanGogh}. It can be seen that in most cases Stable Diffusion does a better job of imitating Van Gogh style than Flux and F-Lite. The performance of Stable Diffusion appears to rely on certain artistic clichés. For example, when recreating paintings by Van Gogh that depict skies, it often reproduces the swirling patterns from his famous Starry Night (see Figure \ref{fig:starry}), even when such elements are absent from the reference original images provided (see Figure \ref{fig:starrySD}). A similar pattern is observed in some of Dalí’s recreations by Stable Diffusion, where clocks appear even when they are absent from the original works\footnote{see works 5, 89, and 389 by Salvador Dalí in \url{https://ama2210.github.io/WikiArt_VLM_Web/}}.

In summary, the accuracy of VLMs in identifying AI-generated images as not being made by human painters depends largely on the text-to-image generator model. For Flux and F-Lite, several VLMs are capable of performing the identification accurately, while for Stable Diffusion, the results are worse. As for the painters, in some cases there seems to be a correlation between performance and artist popularity, but in others there is no such effect. There are also specific artists that tend to get extreme performance values. An example is M.C. Escher which is the worst for real painters and among the best for AI generated images. In this case, it may be due to its particular style that is not recognized either in the original paintings nor in the paintings. AI generated images.

\subsection{Limitations}

The study presented in this paper has several limitations. First, although the dataset of canvas used is extensive, the experimental evaluation can always be extended with additional artists or paintings. The same reasoning applies to both the VLMs and the text-to-image AI models, additional models can be evaluated, in fact by the time the paper is published there will be newer VLMs and text-to-image models. To mitigate this issue, the code and data used in our experiments have been designed to facilitate the testing of new models, for example by releasing the descriptions of all the paintings of the dataset so that they can be used with newer text-to-image models to generate AI imitations. 

Beyond the dataset and models, there are also limitations in the prompts used which target a given artist giving the VLM only the yes or no options for answering. It would be interesting to use open questions on the artist to better understand if the models are capable of identifying the artist or if they attribute the painting to a different one. Similarly, additional analysis in which the models are asked about painters that have similar styles or features in their canvas would also be of interest, as well as doing a finer analysis of the results per painter, genre, and style. To mitigate this issue, data obtained in the evaluation are publicly available for other researchers that can conduct additional analysis.

\subsection{Analysis and discussion}

The results presented in the previous subsections show the limitations of current VLMs to: 

\begin{enumerate}
    \item Identify real paintings that correspond to the original artist.
    \item Identify real paintings that do not correspond to a random artist.
    \item Identify AI-generated images that imitate the style of a painter for some AI text-to-image generation models. 
\end{enumerate}

Only half of the VLMs can reliably identify the content generated by two of the three text-to-image generators used as not corresponding to human painters. Instead, for Stable Diffusion, the detection is not reliable. 

The results also show large variations depending on the artist, with no clear correlation between performance for real and AI-generated images. For some of the AI generators there seems to be a correlation between the artist popularity and the model performance, while that is not the case for real painters.  

Further analysis of the data may provide additional insight into how the performance of VLMs depends on different characteristics of the artist or the painting, such as the style, the genre, or the painting techniques used. These analyses are left for future work, and to facilitate further research, the data is released both in raw format and with an interactive visualization tool. The same applies to the study of the correlation of VLM performance with artist popularity, production, or number of imitations and presence in, for example, merchandising such as mugs, t-shirts or low-cost reproductions \cite{schroeder2006aesthetics}. As mentioned earlier, current systems fail to recognize even what is arguably the most famous painting in the world: La Gioconda.

The limitations of VLMs to perform artist attribution pose a significant risk that can lead to confusion or even misinformation. For example, as users increasingly rely on AI models to answer queries, incorrect information can propagate to millions of users given the widespread adoption of VLMs. However, this is not the only issue, as AI models are also used to process data massively, incorrect information may propagate to websites or other sources of content. For example, VLMs can be used to automatically annotate a large set of paintings that are subsequently published online. In fact, the ease of massive processing of data may be a larger issue than user queries.

To address these issues, ideally the performance of VLMs would improve to reach accuracy values that provide reliable information. However, while that is not the case, VLMs should be carefully used for painting attribution, only as another tool that provides information to take a decision, but not blindly. A good policy for VLMs would be to include warnings or disclaimers in their responses to prevent misinterpretation or misuse of their responses. Another possibility could be to fine-tune the models on a large dataset of paintings to see if the performance improves. A further step would be to include the datasets generated in this work as training data for future VLMs. Both ideas are left for future work, and facilitated by making our datasets public.

\section{Conclusion}
\label{conclusion}

This paper has presented a comprehensive evaluation of state-of-the-art Vision Language Models (VLMs) on tasks: artist attribution for real paintings and detection of AI-generated imitations. Using nearly 40,000 paintings from 128 artists, together with synthetic images generated in the style of those artists, we have demonstrated that most VLMs suffer from substantial limitations in both domains.

First, when attribution of real paintings is made, the best performing VLMs, Gemma3--12B and LLaMa3.2--11B, achieve modest normalized accuracy, while others like GPT4.1-mini and Pixtral-12B show consistent failures or unreliable behavior. Second, when confronted with AI-generated images mimicking painters’ styles, models again vary widely: GPT4.1-mini excels at rejecting attribution, whereas Pixtral-12B often mistakenly credits the suggested artist but results depend heavily on the AI generator used to create the images. 

These findings expose important risks: as users increasingly rely on VLMs for artist information, errors may lead to widespread confusion or misinformation. The potential scale of harm increases as AI annotations proliferate online and across downstream applications. To mitigate these risks, we recommend caution in the deployment of VLM-based attribution tools using them as decision-support tools rather than definitive authorities.


\section*{Acknowledgments}
This work is supported by the FUN4DATE (PID2022-136684OB-C22) and SMARTY (PCI2024-153434) projects funded by the Spanish Agencia Estatal de Investigación (AEI) 10.13039/501100011033, by TUCAN6-CM (TEC-2024/COM-460), funded by CM (ORDEN 5696/2024) and by the Chips Act Joint Undertaking project SMARTY (Grant no. 101140087).

\bibliographystyle{unsrt}  
\bibliography{sample-base}

\end{document}